\documentclass[12pt]{iopart}

\usepackage[english]{babel}

\usepackage[utf8]{inputenc}

\usepackage{makecell}

\usepackage{verbatim}

\usepackage{natbib}

\usepackage{epsfig}

\usepackage{graphicx}

\usepackage{url}

\usepackage[hidelinks]{hyperref}

\newcommand{\bibtit}[1]{\textit{#1}}

\newcommand{\bibjou}[1]{\textit{#1}}

\newcommand{\bibjtit}[1]{{#1}}

\newcommand{\bibaut}[1]{{#1}}

\newcommand{\biburl}[1]{{\footnotesize\url{#1}}}

\newcommand{\rs}{\,r_\mathrm{S}}

\begin{document}

\title[Virtual sector models: \textit{V-SeMo}]{%
\textit{V-SeMo}: a digital learning environment
for teaching general relativity
with sector models}

\author{Sven Weissenborn, Ute Kraus and Corvin Zahn}

\address{Institut für Physik, Universität Hildesheim, Universitätsplatz 1,
  31141 Hildesheim,
  Germany}

\eads{\mailto{weissenborn@uni-hildesheim.de}, \mailto{ute.kraus@uni-hildesheim.de},
  \mailto{corvin.zahn@uni-hildesheim.de}}

\vspace{10pt}
\begin{indented}
\item[]May 28, 2024
\end{indented}

%---------------------Abstract-----------------------------
\begin{abstract}
The teaching of general relativity
at the secondary school and lower undergraduate university levels
is necessarily based on approaches
with a restricted use of mathematics.
An important aspect of teaching general relativity at this level
are suitable learner activities.
While a substantial number of such activities
has been reported
for studying the non-Euclidean geometry of curved surfaces,
there are far fewer reports of activities
that let learners actually study spacetimes
and infer physical phenomena from their geometry.
In this article we report on the digital learning environment
\textit{V-SeMo} that brings sector models
(\citealt{zahn2014})
into an interactive web application.
\textit{V-SeMo} lets learners explore relativistic spacetimes
by constructing geodesics and assessing curvature.
We describe the didactic design and the user interface
of \textit{V-SeMo},
discuss the extended possibilities of virtual sector models
compared to the paper models described previously,
and present an activity on light deflection near neutron stars
by way of example.
We further report on an evaluation carried out
with secondary school students.
Results indicate a high learning effectiveness in both
the \textit{V-SeMo}-based and
the paper-based versions of a teaching unit
on relativistic light deflection.
The teaching materials presented in this article are available
online for teaching purposes at https://www.spacetimetravel.org.
\end{abstract}
%----------------------------------------------------------

\vspace{2pc}
\noindent{\it Keywords\/}: general relativity,
digital learning environments,
secondary school education,
undergraduate university education

\newpage

%----------------------------------------------------------

\section{Introduction}

To provide insight into
today's physical world view
counts among
the goals of teaching physics at all levels.
General relativity makes an important contribution to this goal,
being the theory that describes
our physical understanding of space and time,
and of gravity in terms of spacetime geometry.
There is a broad class of learners of general relativity
who
profit from approaches
that do not go much beyond secondary-school-level mathematics,
and are
based on the use of models and simulations.
These learners include
secondary school students
and also some groups of university students,
e.g. pre-service physics teachers,
and students with a focus on experimental physics.
The development of new approaches,
curricula and learning resources
for general relativity
designated for this group of learners
is a topical area of research in physics education.

For physics teaching in a formal setting
such as school or university,
learner activities and exercises
play a central role.
They let learners develop knowledge and skills,
gain practice, deepen their understanding,
and apply what they have learned to new situations.
Hence,
general relativity courses
of the type described above
require
appropriate
learner activities
that are tailored to
their mathematically restricted approaches.

In teaching the modern, geometrical view
of gravity,
key parts of any course are
an introduction to non-Euclidean geometry,
and
the application of geometric concepts
to relativistic spacetimes.
For the introduction to non-Euclidean geometry,
a substantial number of
activities at the
level considered here
has been reported.
They let learners study the geometry of curved
surfaces,
providing hands-on tasks
that utilise physical models,
measurement and geometric construction.
Many make use of readily
available objects from everyday life.
Tasks have been proposed
that involve the hands-on construction of geodesics,
e.g. by taping strips of paper to the surface in question,
and the use of geodesics to study
the angle-sum of geodesic triangles,
the circumference-radius relations of circles,
or the run of initially parallel geodesics
(\citealt{eff1977}, \citealt{wood2016}, \citealt{kaur2017},
\citealt{gasp2018}, \citealt{taim2018},
\citealt{ryst2019, ryst2021, ryst2022},
\citealt{kraus2021}).

However, far fewer activities have been reported
for learners to study the geometry
of specific relativistic spacetimes
with methods that are accessible to
the groups of learners considered here.
These activities have in common
that they use purpose-built models
representing two- or three-dimensional subspaces
of the spacetime to be studied.
The models are designed in such a way
that hands-on methods
or calculations on the level of secondary school mathematics
suffice to determine geodesics or curvature.
\cite{ryst2021} considers the Schwarzschild spacetime
and describes the hands-on construction of
geodesics in the equatorial plane
making use of a 3D-printed model of Flamm's paraboloid.
\cite{dis1981} describes an algorithm
for the numerical tracing of geodesics;
he proposes that learners implement this algorithm
and then study geodesics with a computer simulation.
The use of so-called wedge maps,
a class of discrete models
for geometries with a centre of symmetry,
makes it possible
to formulate the algorithm
with secondary school-level mathematics.
Applications are described
for the Schwarzschild spacetime.
Zahn and Kraus (\citealt{zahn2014, zahn2019}, \citealt{kraus2019})
propose activities using hands-on methods
on so-called sector models.
Sector models are
a class of discrete models
for surfaces, spaces, and spacetimes
that permit the tracing of geodesics
with pen and ruler
and the study of curvature in two and in three dimensions.
Zahn and Kraus describe applications for the Schwarzschild spacetime.

For the teaching of general relativity
at the level described above,
it is a desideratum
to have
a larger pool of
learner activities
on relativistic spacetimes
with a greater variety of tasks.

The present work addresses this desideratum.
We report on the development and test of
a new learning tool
in the form of a web application
that lets learners study relativistic spacetimes.
The learning environment \textit{V-SeMo} (Virtual Sector Models)
is based on the
use of sector models
that represent two-dimensional subspaces
of curved spacetimes.
\textit{V-SeMo} provides sector models in virtual form
together with tools that
permit to manipulate the models interactively
and that furthermore
support their use with features including
speed-up by automation, elements of scaffolding,
and guided activities.
Going from paper sector models to virtual sector models,
we widen the scope of
what learners can do with these models
and we include additional features
that support the learning process
and also support instructors in classroom use.
Our \textit{V-SeMo} applications are
accessible online
for teaching use.
Moreover, the \textit{V-SeMo} project is open source.
See the online resources for this article
(\citealt{weiss2024})
for access to the applications and the code.

After a brief introduction to sector models
in section~\ref{sec:sector_models},
we will
in section~\ref{sec:V-SeMo}
describe the user interface,
the features and the didactic design of \textit{V-SeMo}.
An example of a \textit{V-SeMo}-based activity
is given in section~\ref{sec:teaching_example}.
In section~\ref{sec:learning_effectivness}
we describe an evaluation carried out with secondary school students.
Discussion and conclusions follow in section \ref{sec:discussion}.

%----------------------------------------------------------

\section{Sector models}
\label{sec:sector_models}

\begin{figure}
  \centering
  \begin{minipage}[t]{0.45\textwidth}
  \includegraphics[width=\textwidth]{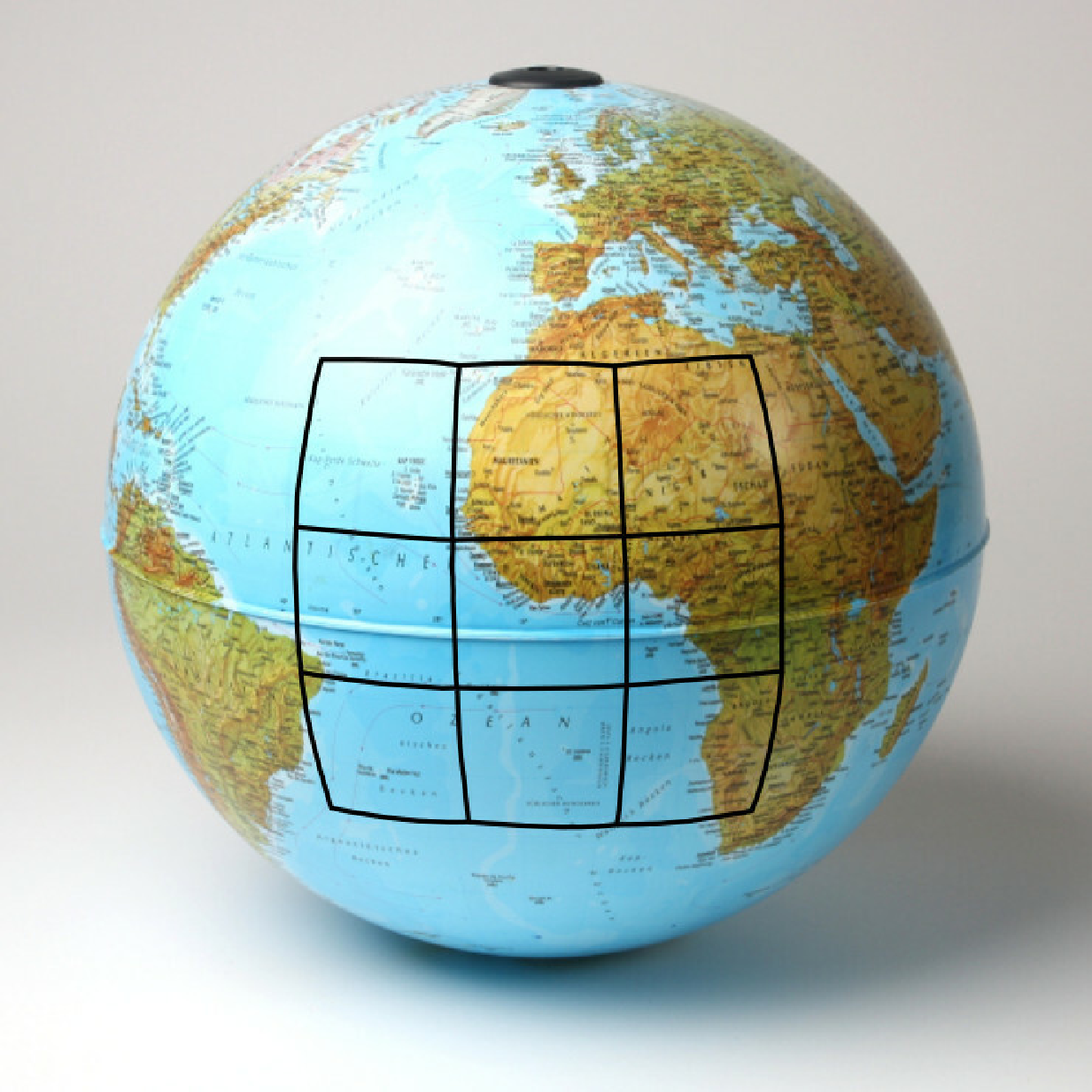}
    (a)
  \end{minipage}
  \quad
  \begin{minipage}[t]{0.45\textwidth}
      \includegraphics[width=\textwidth]{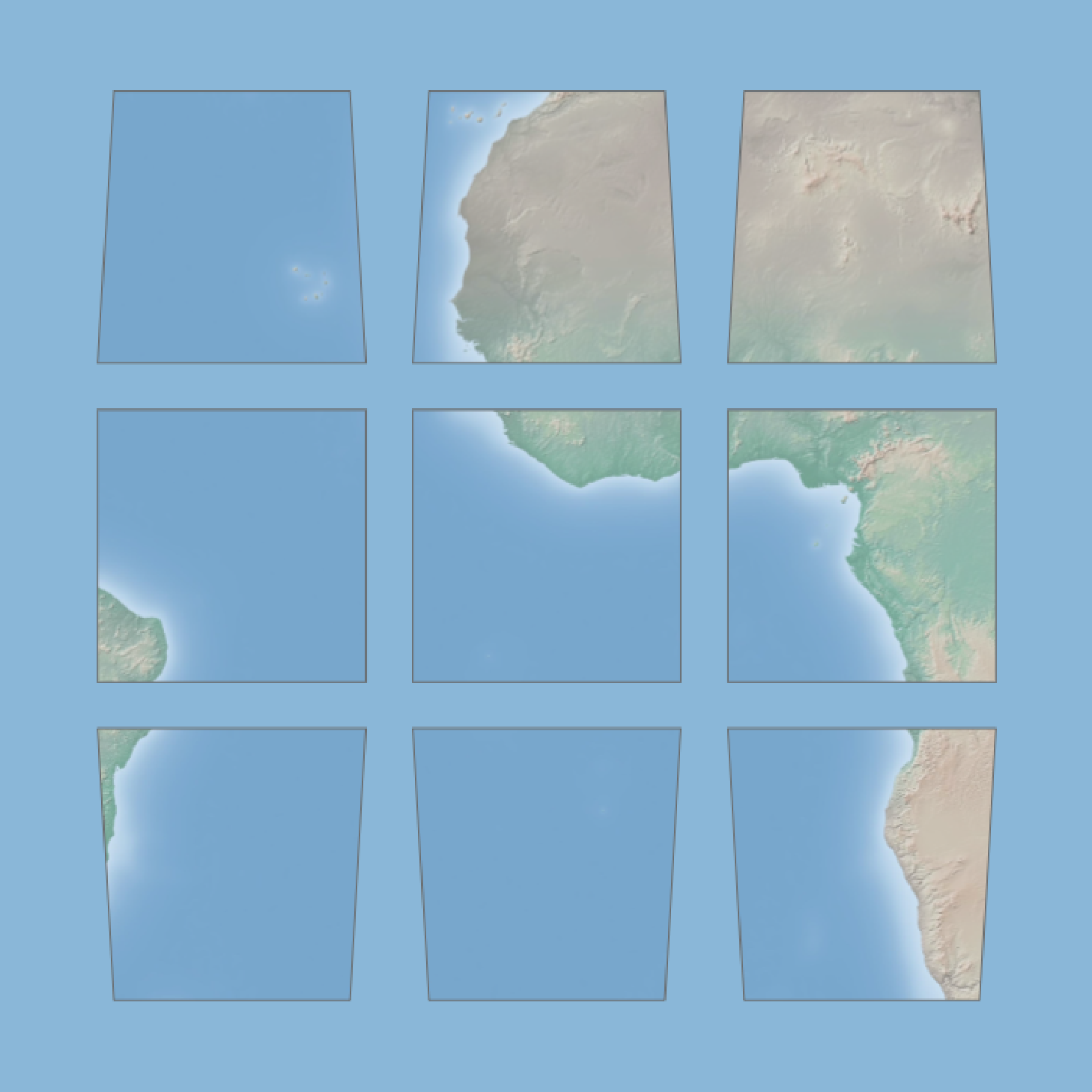}
    (b)
  \end{minipage}
  \caption{Creation of a sector model: Part of a sphere is divided
    into small elements of area (a). These small subareas are then
    approximated by small flat pieces (b).}%
  \label{fig:from_globe_to_sectors}
\end{figure}

We briefly summarise the construction
and use of
sector models in two dimensions
(\citealt{zahn2014, zahn2019}, \citealt{kraus2019}).
These models can
represent two-dimensional spaces
and spacetimes with curvature.
We first consider spatial sector models
and take a curved surface, the sphere,
by way of example.
There are two basic steps involved
in creating a sector model
of the sphere
(figure~\ref{fig:from_globe_to_sectors}):

\begin{enumerate}
\item \textbf{Surface Subdivision}
(figure~\ref{fig:from_globe_to_sectors}(a)):
The surface is divided into small elements of area.
\item \textbf{Flat Approximation}
(figure~\ref{fig:from_globe_to_sectors}(b)):
Each surface element is approximated by a flat piece
with identical edge lengths and symmetry.
The flat pieces are the `sectors'.
\end{enumerate}

Arranged in a planar layout,
these sectors collectively form the sector model
of the curved surface.
They provide a true-to-scale representation
with an accuracy determined by
the fineness of the subdivision.

\begin{figure}
  \centering
  \begin{minipage}[t]{0.45\textwidth}
      \includegraphics[width=\textwidth]{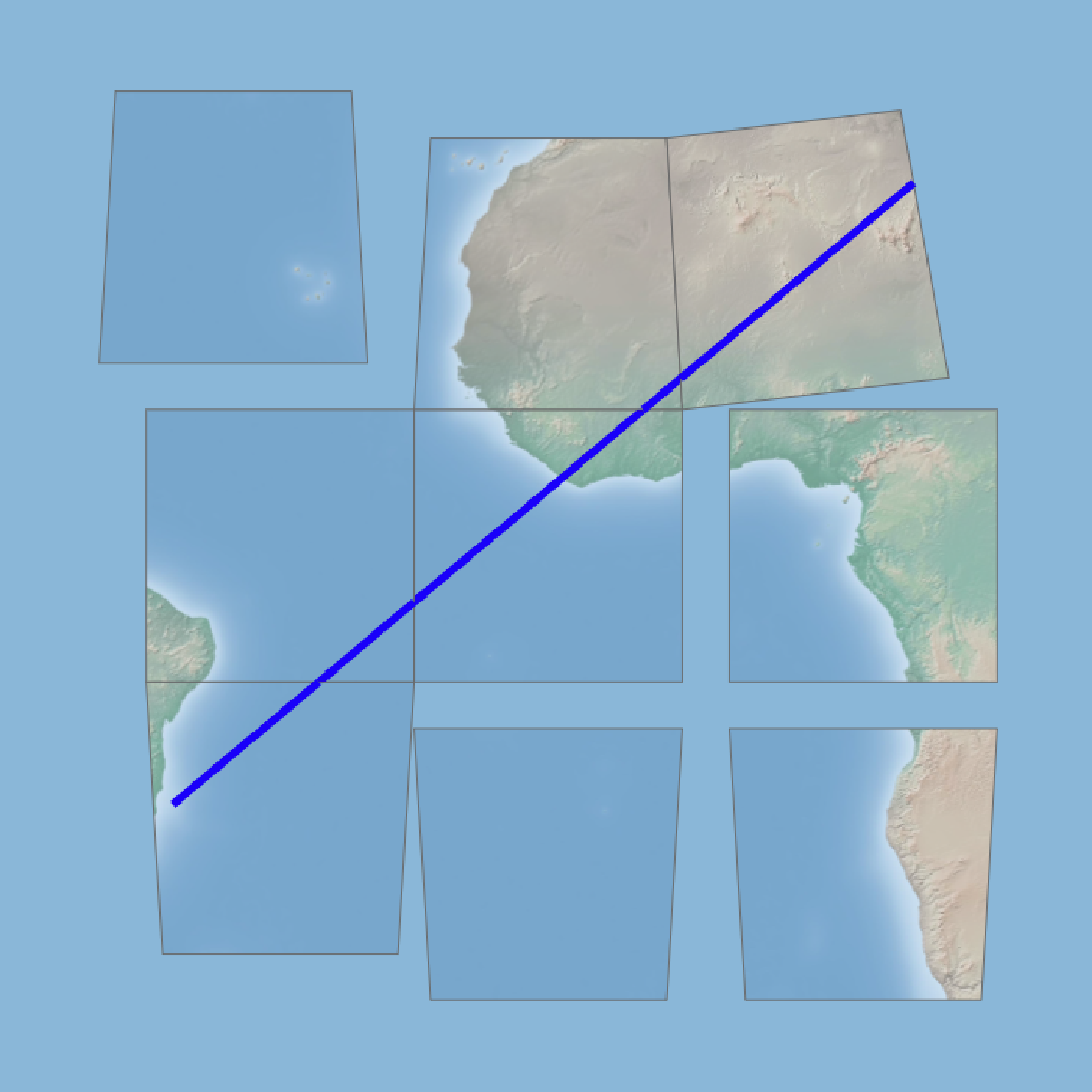}
    (a)
  \end{minipage}
  \quad
  \begin{minipage}[t]{0.45\textwidth}
    \includegraphics[width=\textwidth]{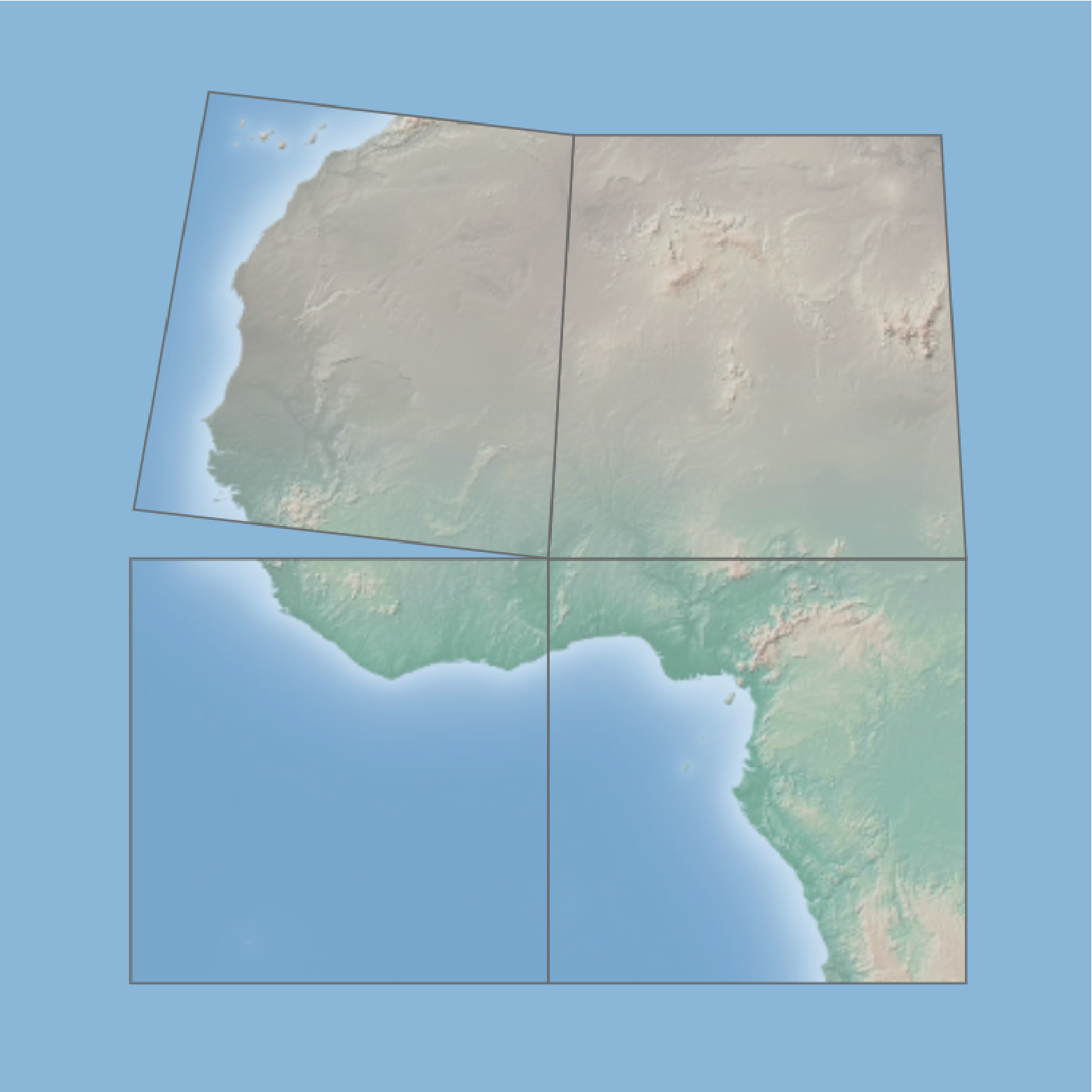}
    (b)
  \end{minipage}
  \caption{Two applications of sector models. (a) Construction of a
    geodesic. (b) Test for curvature.}%
  \label{fig:sectors_in_action}
\end{figure}

Figure~\ref{fig:sectors_in_action}(a)
illustrates the use of the sector model
for finding geodesics.
The procedure follows directly from the definition
of a geodesic as a locally straight line.
Within a sector the geometry is Euclidean,
and a geodesic is a straight line.
When a geodesic reaches a sector edge,
it seamlessly continues in the same direction
into the neighbouring sector.

A second important use of sector models
is the determination of curvature.
Figure~\ref{fig:sectors_in_action}(b)
illustrates a qualitative test:
Four sectors with a common vertex are joined.
If they fit together without a gap,
this indicates a flat surface with zero curvature.
The presence of a gap,
as shown in figure~\ref{fig:sectors_in_action}(b),
signifies positive curvature,
while overlapping sectors indicate negative curvature.

\begin{figure}
  \centering
  \begin{minipage}[t]{0.45\textwidth}
    \includegraphics[width=\textwidth]{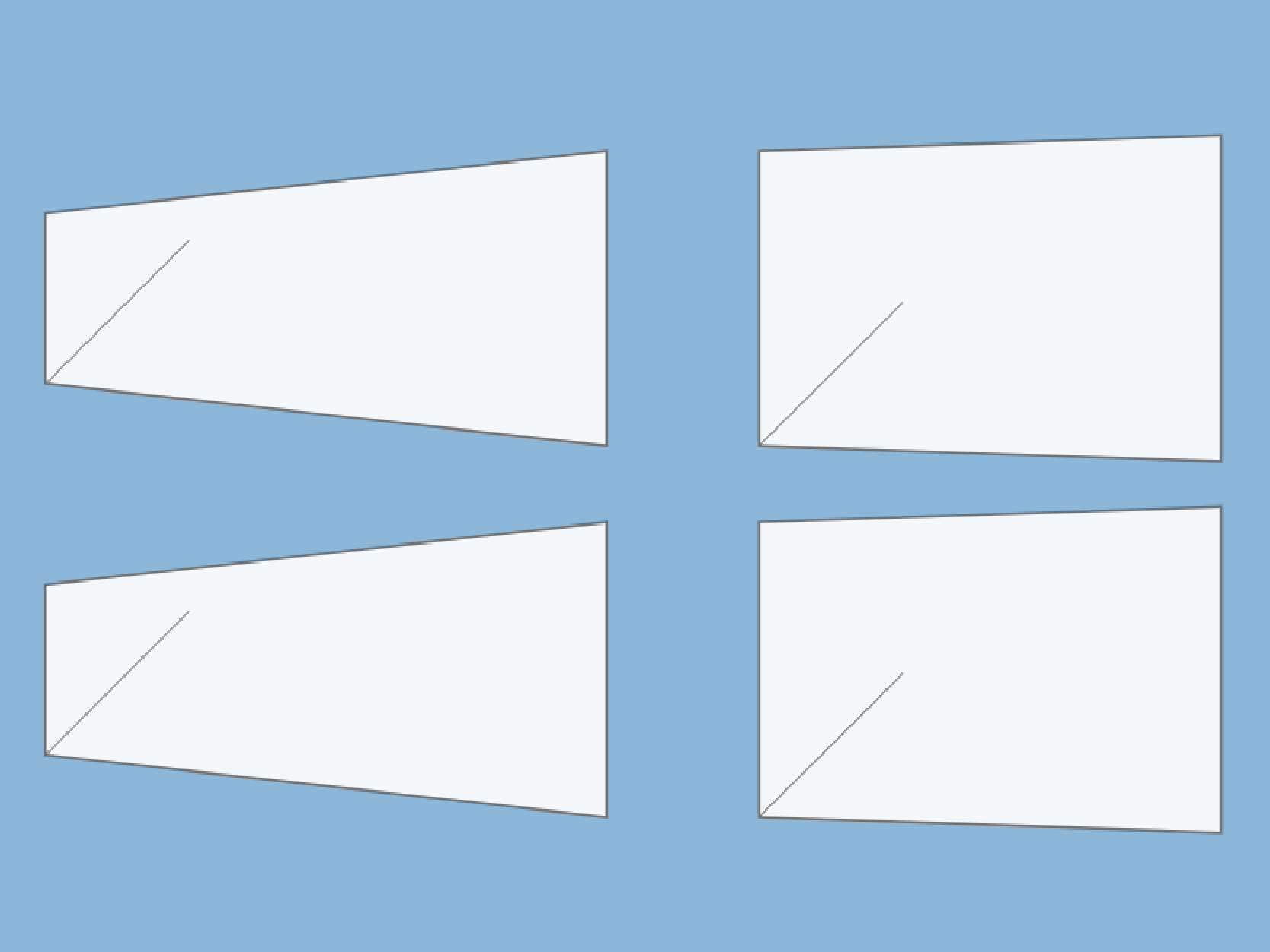}
    (a)
  \end{minipage}
  \quad
  \begin{minipage}[t]{0.45\textwidth}
    \includegraphics[width=\textwidth]{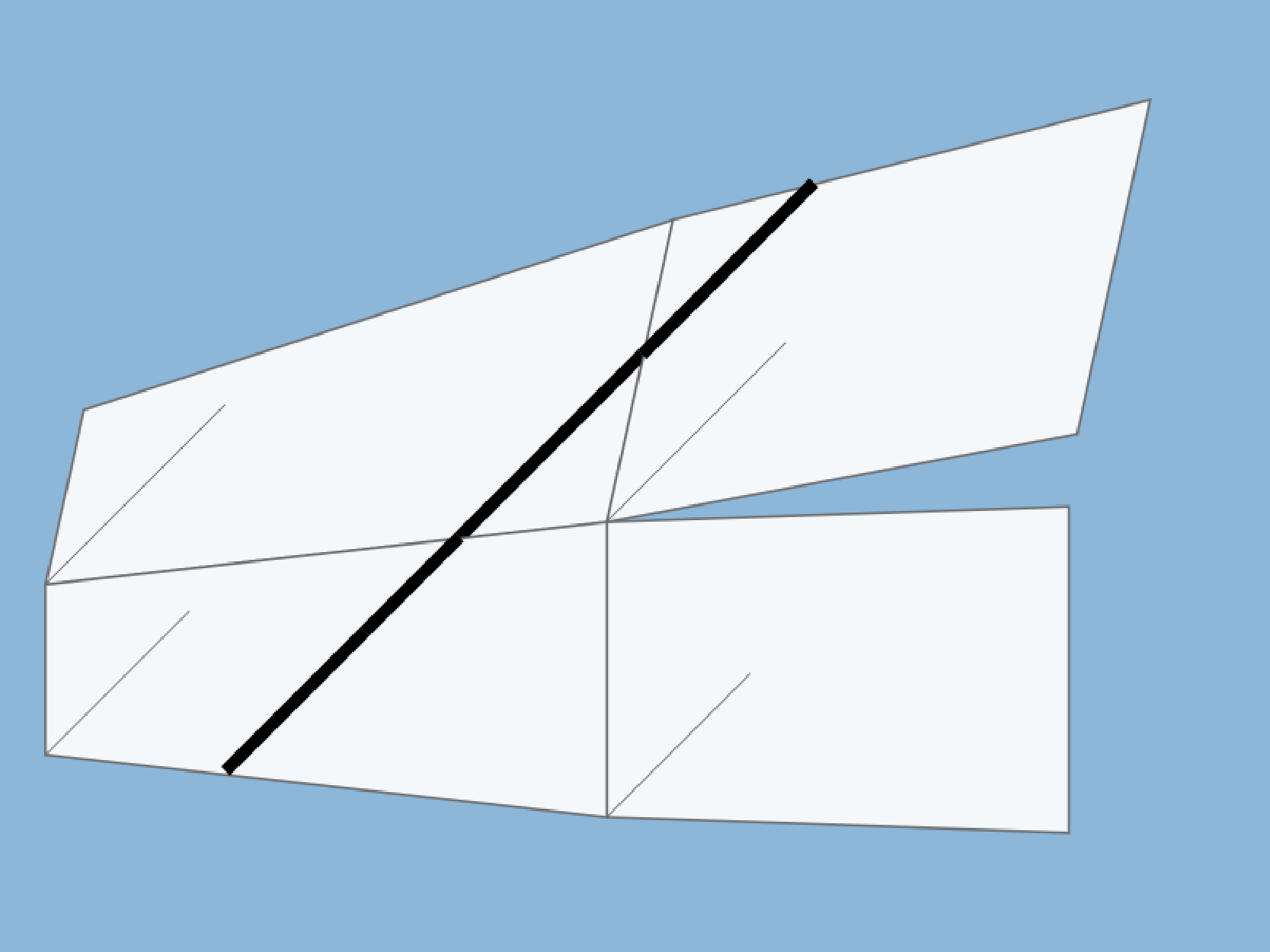}
    (b)
  \end{minipage}
  \caption{Spacetime sector model representing a radial ray in
    Schwarzschild spacetime (\citealt{zahn2014}), the diagonal thin
    lines are light cones (a). Construction of a radial spacetime
    geodesic and test for curvature (b).}%
  \label{fig:spacetime}
\end{figure}

The creation and use of sector models
for spacetimes is analogous to the spatial case.
We here consider spacetimes restricted to one spatial direction
(1+1-dimensional spacetimes),
and illustrate the procedure
for the case of a radial ray
in the exterior Schwarzschild spacetime
(figure~\ref{fig:spacetime}).
After subdivision of the spacetime,
each element of spacetime
is approximated by a piece of Minkowski spacetime
with the same edge intervals and symmetry,
these are the spacetime sectors
(figure~\ref{fig:spacetime}(a)).
Constructing geodesics and
determining curvature
proceeds as in the spatial case
(figure~\ref{fig:spacetime}(b)).
Note that joining spacetime sectors
involves Lorentz transformations
for aligning the edges of adjacent sectors,
in lieu of the rotation applied to spatial sectors.

Determining curvature by way of calculation
requires computing the Riemann curvature tensor,
and geodesics
are found
by solving a system
of ordinary differential equations.
These mathematical techniques
are beyond the scope of school education
and in part also of undergraduate education
as argued above.
With the use of sector models,
computation is replaced
with
geometric construction,
making the geometry of curved spaces
and spacetimes
accessible to a broader group of learners.

In the
workshops described
in previous work
the two-dimensional sector models
were made from paper
(\citealt{zahn2019}, \citealt{kraus2019}).
They were either built as models with movable sectors,
by cutting individual sectors out of sheets of paper
and gluing them onto cardboard with
repositionable spray adhesive.
Or models with fixed sector positions
were printed on worksheets,
and used with a small number of
cut out so-called transfer sectors.

%----------------------------------------------------------

\section{The digital learning environment \textit{V-SeMo}}
\label{sec:V-SeMo}

\begin{figure}
  \centering
  \begin{minipage}[t]{0.45\textwidth}
  \includegraphics[width=\textwidth]{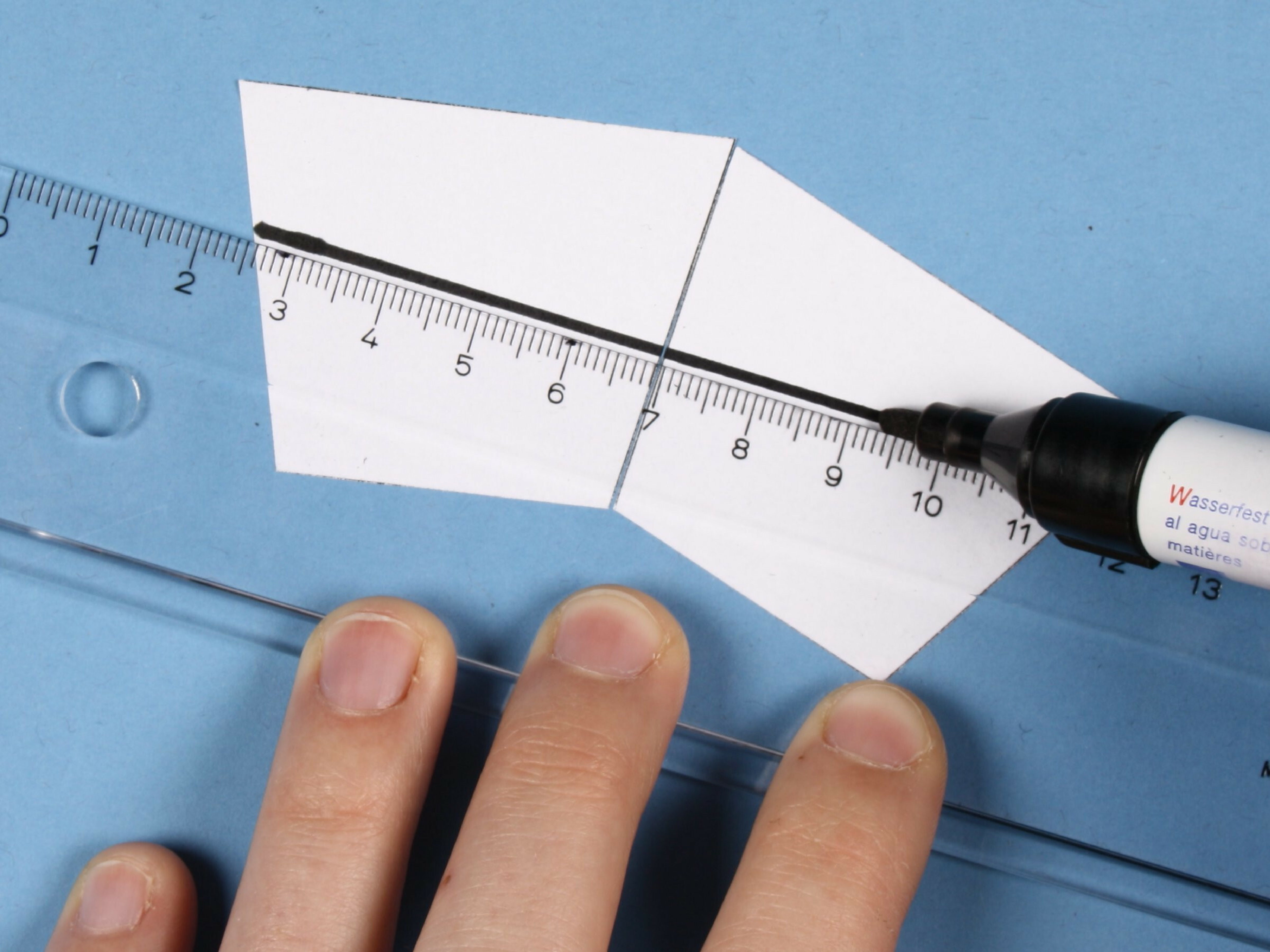}
    (a)
  \end{minipage}
  \quad
  \begin{minipage}[t]{0.45\textwidth}
    \includegraphics[width=\textwidth]{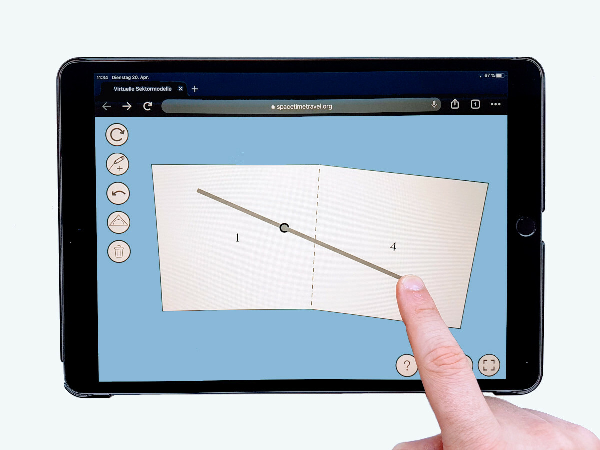}
    (b)
  \end{minipage}
  \caption{The construction of a geodesic on a paper model
    (\citealt{zahn2019}) (a) and on a virtual sector model (b).}%
  \label{fig:draw_geodesics}
\end{figure}

Here we report on the development of
a digital representation
of sector models and their handling in the digital learning
environment \textit{V-SeMo}.
Figure~\ref{fig:draw_geodesics}
illustrates the move from paper sector models to
virtual sector models using the construction of a geodesic
by way of example:
With a paper sector model, paper sectors are placed
next to each other,
and a geodesic is then drawn with pen and ruler
(figure~\ref{fig:draw_geodesics}(a)).
With a virtual sector model,
the joining of sectors and the drawing of lines
proceeds with drag gestures
(figure~\ref{fig:draw_geodesics}(b)).
The learning environment
\textit{V-SeMo} is an interactive web application,
accessible through standard web browsers without prior installation.
For ease of use in educational contexts,
\textit{V-SeMo} is designed to run in web browsers
both on computers and on tablets.
This online environment provides
all hitherto published 2D and 1+1D sector models
and so facilitates the implementation of the
associated workshops described previously
(\citealt{zahn2014, zahn2019}, \citealt{kraus2016, kraus2019}, \citealt{kraus2021}, \citealt{weiss2022}).
It also provides the guided activity
described below;
more content will be added in the future.

\subsection{Basic functions and operating design}

\begin{figure}
  \centering
  \includegraphics[width=\textwidth]{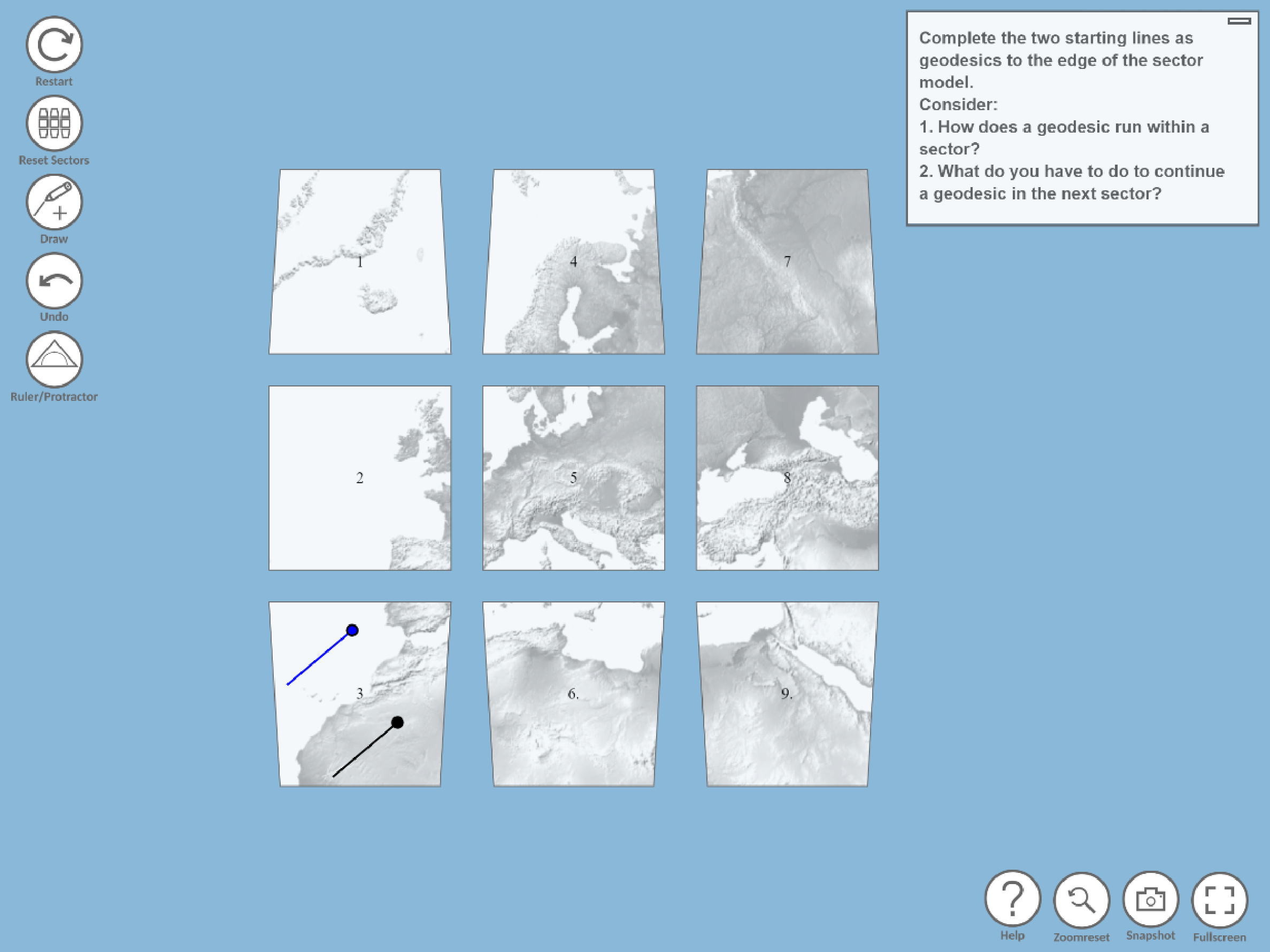}
  \caption{A screenshot of the interface of \textit{V-SeMo}, showing
    the main workspace with the display of a sector model (centre),
    the basic toolbars (left and bottom), and the information panel
    (top right).}%
  \label{fig:V-SeMo}
\end{figure}

\textit{V-SeMo} has been designed
with the aim to make its use intuitive
so that users can quickly focus on working with the models
rather than on learning to operate the application.
To this end,
\textit{V-SeMo} was developed
taking
research results on intuitive interface design
into account
(\citealt{adams2008}),
and development was
an iterative process
involving usability tests with both individuals
and groups, including secondary school students,
university students and teachers.
The iterative design process
shaped
the user interface,
information presentation,
and overall interaction design.

The fundamental operations required for
handling sector models
that \textit{V-SeMo} has to provide are:
moving sectors, rotating/transforming sectors,
joining sectors, and drawing lines.
Figure~\ref{fig:V-SeMo} shows a screenshot of
the \textit{V-SeMo} interface.
\textit{V-SeMo} offers a digital workspace at its centre,
much like drawing software.
Most interactions occur in this workspace
that occupies the major part of the display.
Users can zoom in or out
and navigate the workspace
using standard mouse or touchscreen gestures.
Toolbars frame the workspace
and provide the
required
functions.
In the left toolbar,
shown in figure~\ref{fig:V-SeMo}
in its basic configuration,
`Restart' resets the entire application,
while `Reset Sectors'
restores all sectors to their initial positions
(keeping lines drawn in the sectors unchanged).
`Draw' changes into drawing mode,
`Undo' reverts a single operation,
and
`Ruler/Protractor' provides a Geodreieck
(a set square with integrated protractor)
in the workspace.
More tools can be added to the left toolbar
as described below.
The bottom toolbar provides help and service functions.

\textit{V-SeMo} operates in one of three distinct modes:
grabbing objects, drawing lines, and manipulating geodesics.

\begin{figure}
  \centering
  \begin{minipage}[t]{0.45\textwidth}
    \includegraphics[width=\textwidth]{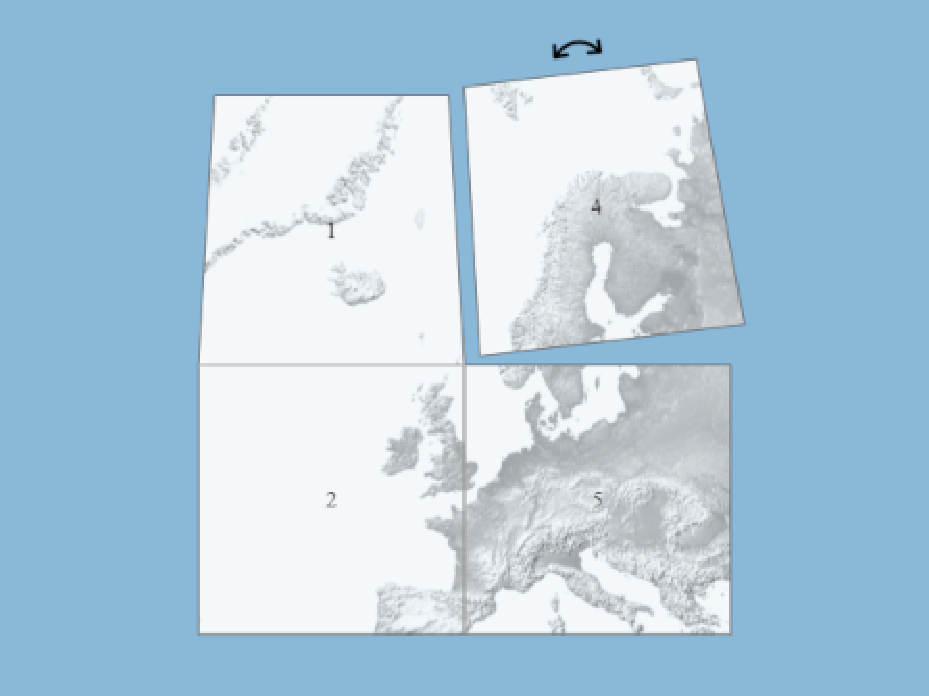}
    (a)
		\vspace{0.2cm}
  \end{minipage}
  \quad
  \begin{minipage}[t]{0.45\textwidth}
    \includegraphics[width=\textwidth]{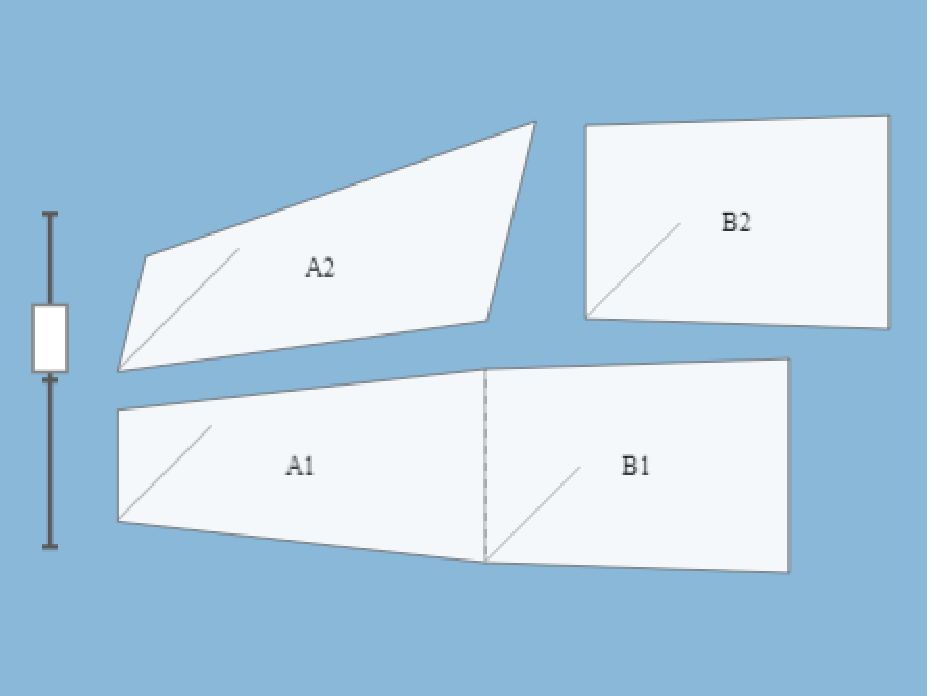}
    (b)
    \vspace{0.2cm}
  \end{minipage}
  \caption{Spatial sectors can be rotated using a rotation arrow (a),
    spacetime sectors can be Lorentz-transformed using a rapidity
    slider, here shown for sector A2 (b).}%
  \label{fig:manipulate_sectors}
\end{figure}

In the initial state, the `Grab' mode is active,
allowing users to move objects with the mouse or a finger.
Spatial sectors can be rotated using the usual rotation gesture
or a displayed rotation arrow
(figure \ref{fig:manipulate_sectors}(a)).
Spacetime sectors
undergo shape changes via Lorentz transformation.
The velocity for a Lorentz transformation
is adjusted using a sector-specific slider
that appears after clicking or touching
a spacetime sector
(figure~\ref{fig:manipulate_sectors}(b)).
The slider position corresponds to
the rapidity $\alpha$
(velocity $v$
being given by $v/c = \tanh\alpha$).
All line segments in a sector
are rotated or Lorentz transformed
together with the sector.

In `Draw' mode,
users can create new lines in sectors
or extend existing ones.
To draw a new line, they select the `Draw' button
and drag the end point of the line across sectors.
Extending a line is possible from its endpoint,
and each line has a drag point for this purpose,
eliminating the need to press the `Draw' button.

The `Manipulate Geodesics' mode
offers additional functions
accessible through
buttons in the left toolbar
that only appear after clicking or touching a line.
This mode allows users to delete an entire line
or manipulate it as a geodesic.
Details about the `Manipulate Geodesics' mode
and its functions are given
in section \ref{sec:automanipulate}.

The user interface remains consistent
across all sector models,
making it easy for learners to adapt to new models.
It is possible to limit the set of available
functions initially,
for clarity, and to expand it as
learners progress.

\subsection{Didactical design elements}

To facilitate effective learning of the correct handling
of sector models,
\textit{V-SeMo} incorporates various didactic design elements.
They are designed to support independent learning,
and they address,
in particular, the following two
fundamental aspects of using sector models:

  \begin{enumerate}
    \item   By construction, each sector has
            well-defined neighbours.
            Much like assembling a puzzle,
            only sectors that are neighbours
            may be joined at their common edge.
    \item   When a geodesic reaches
            the edge of a sector,
            the correct neighbouring
            sector must be appended
            before the geodesic can be
            extended across the sector edge.
  \end{enumerate}

During our initial testing phase,
a number of difficulties were identified
that learners had with these aspects of sector models.
The most common errors
were the incorrect joining of sectors
and the drawing of lines on the background,
i.e., outside of the sectors.
To tackle these difficulties,
we incorporated elements of
`implicit scaffolding' (\citealt{podo2010})
into the design of \textit{V-SeMo}.
This
as described by \cite{paul2013} implies:
  \begin{quote}
    ``Implicit scaffolding is meant
    to allow for student autonomy, the feeling
    that students have independent control
    over their experience, while both
    affording and constraining students
    to actions that are productive for
    learning.''
  \end{quote}
Implicit scaffolding aims at integrating guidance
seamlessly into the design of the application,
guiding the learning process without explicit instruction.
Key components
include visual cues and
program-controlled support or constraints.

An example for a visual cue in \textit{V-SeMo} is given in
figure~\ref{fig:V-SeMo}:
The sector model of the sphere is provided with a texture
in the form of a world map.
The connections between the sectors
are apparent from the map:
Only the correct neighbours reproduce the world map correctly.
This supports the correct joining of sectors
and so addresses point (i).
\textit{V-SeMo} also provides program-controlled support for
sector joining.
When learners position a sector near a valid neighbour,
\textit{V-SeMo} aligns and snaps it to the target.
The successful assembling of the sectors is confirmed
by a colour change of their common edge line.
This support of sector pairing
is also provided in the case of spacetime sectors.
Here, aligning sectors usually requires a Lorentz transformation.
When the user
places a sector sufficiently close
to a valid neighbour,
the application automatically joins the sectors,
choosing the required position
and Lorentz transformation.
The common edge is marked in colour as in the spatial case.

\textit{V-SeMo} also supports the correct drawing of geodesics:
Drawing on the background
or on overlapping sectors
is prevented.
If the user
attempts such an incorrect drawing action,
\textit{V-SeMo} truncates
the line at the edge of the sector.
A geodesic can only be extended beyond a sector edge
when the correct neighbour has been appended.
Other features of the application are provided
to make the construction of geodesics easier:
The end point of each line is marked
and the user
can draw a new line segment
connected to an existing line.
If the direction is approximately the same as the previous line,
\textit{V-SeMo} adjusts it automatically to agree exactly.

\subsection{Automation and geodesic manipulation}
\label{sec:automanipulate}

\begin{figure}
  \centering
  \includegraphics[width=\textwidth]{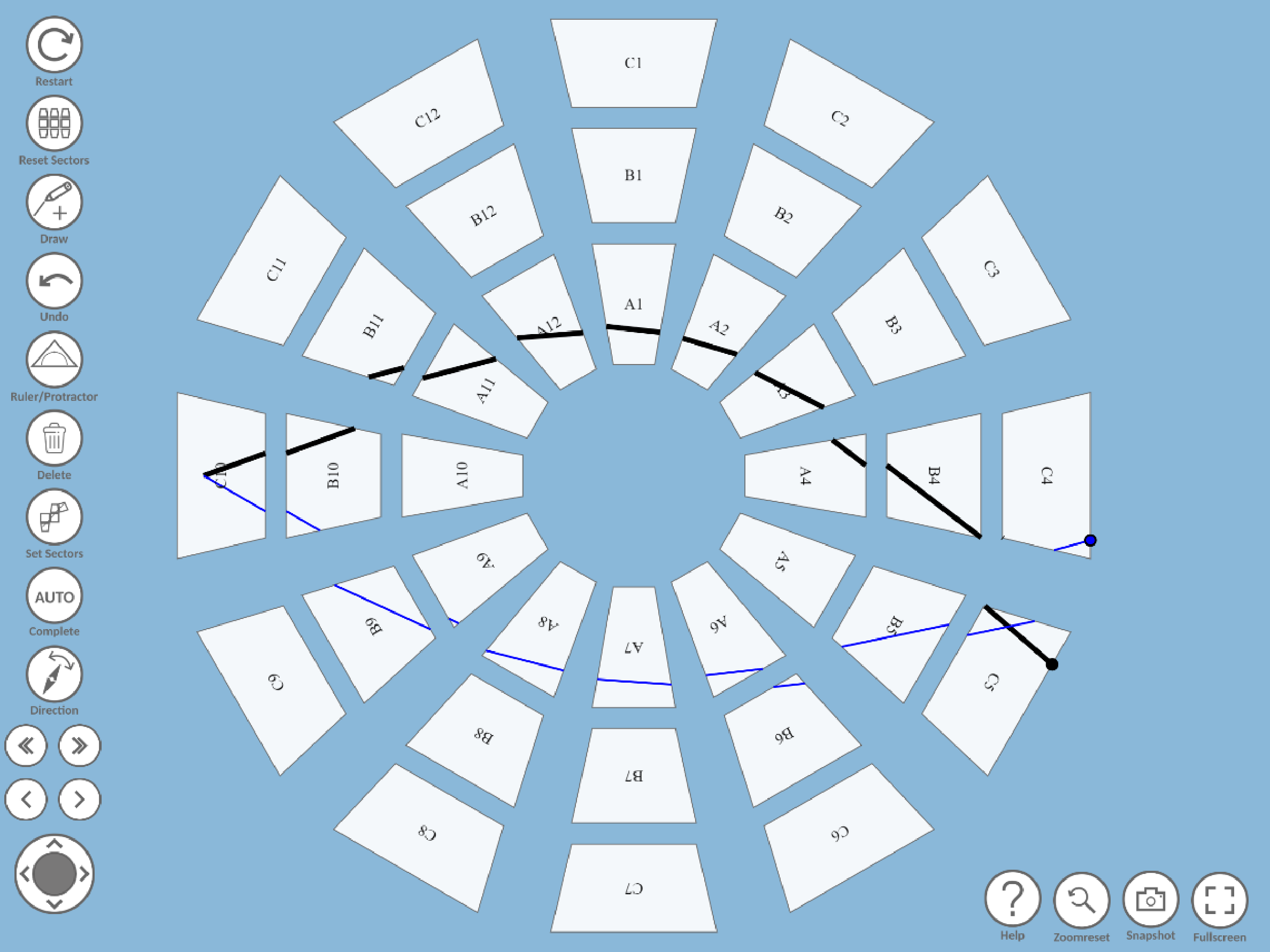}
  \caption{A screenshot of the interface of \textit{V-SeMo} in
    `Manipulate Geodesics' mode. The sector model represents the
    equatorial plane of a black hole (\citealt{zahn2019}). Both lines
    are geodesics. The upper line was selected and is displayed with
    increased line width. Additional buttons appear in the left
    toolbar.}%
  \label{fig:V-SeMo_bl}
\end{figure}

In its `Manipulate Geodesics' mode,
\textit{V-SeMo} provides automated features
that speed up the construction of geodesics.
They are recommended for use by learners
who are already familiar
with the concept of sector models
and proficient in their handling
(including in particular the aspects (i)
and (ii) described above).
To enter this mode,
users select a line,
by either touching it or by clicking on it.
Once a line is selected,
this line becomes visibly thicker
and additional buttons appear in the left toolbar
as shown in figure~\ref{fig:V-SeMo_bl}.
\textit{V-SeMo} also provides the possibility
to enable these buttons individually
via URL parameters\footnote{%
A guide to the use of URL parameters
in \textit{V-SeMo} is provided
in the online resources for this article
(\citealt{weiss2024}).
}.

The extended toolbar provides
line editing functions
for geodesics
that streamline the process of constructing them:
`Set Sectors' identifies all the sectors
crossed by the selected geodesic and its extension,
and it displays them joined in a row.
This allows the user
to quickly
extend the geodesic in one go,
replacing the step-by-step method of joining the next sector
one at a time and extending the geodesic across it.
The `Complete' function takes automation one step further
and automatically completes the selected geodesic,
leaving the position and orientation of the sectors unchanged.
`Delete' removes the entire selected line.

Using the `Direction' function
\textit{V-SeMo} provides dynamic control over the starting direction
of the geodesic.
A click on `right arrow'
turns the starting direction
clockwise by 0.1 degrees,
a click on `double right arrow'
does so by 1 degree,
while `left arrow' and `double left arrow'
do the same counterclockwise.
Moving the joystick
produces a shift of the starting point
in the corresponding direction.
After each adjustment,
of starting direction or starting point,
\textit{V-SeMo} automatically completes the geodesic.

These automation and manipulation features
significantly simplify the construction of geodesics.
Being much faster than step-by-step construction,
they allow to quickly study long geodesics,
or a larger number of geodesics,
and they make geodesic construction on large sector models
manageable.
Also, learners
can use them to easily study
the impact of the start parameters on a geodesic.
In sum, the features of geodesic manipulation
allow the proficient user
to focus on the properties of geodesics
and their physical significance
rather than on the process of constructing them.

In section \ref{sec:teaching_example}
we present a learner activity
that heavily uses the
automation and manipulation functions
described here
and indeed would not be viable without them.

\subsection{Going from Paper-Based to Virtual Sector Models}

In this section we compare
paper-based and virtual sector models
with regard to their use in the classroom.

An important practical aspect is the handling
of sector models:
the preparation of materials,
the ease of use
and the time required to fulfil tasks.
For the comparison in handling
we draw on experience gained
in workshops using one or the other representation.
The effort required to prepare the sector models
varies according to the chosen form of presentation.
If paper models are used, they need to be prepared as cut-out sheets
and cut out
accurately by the learners.
The number of sectors required will influence
the effort required to prepare them.
If virtual models are used instead of paper models,
the effort is significantly reduced
because they
can be accessed online
in the browser.
This requires
technical infrastructure,
digital devices,
and
internet access.
The test for curvature described
in section~\ref{sec:sector_models}
is identical for paper-based and virtual sector models:
the sectors must be placed around a common corner point.
In terms of the time required to perform the tests,
the difference is marginal.

The manual procedure of constructing geodesics
on sector models
is similar
for both types of representation.
To continue a given starting line as a geodesic
on the sector model,
the sectors must be placed adjacent to each other
and the line must be continued in a straight direction.
Unlike the paper model,
the sectors of the virtual model cannot accidentally slip.
There is also no need for a ruler,
as \textit{V-SeMo} enforces the straight geodesic course.
This makes manual geodesic construction on virtual models faster.
In addition, the implemented scaffolding elements
support the learner in joining
sectors
and prevent drawing over sector edges
where there are no valid neighbours.
If nevertheless mistakes are made during drawing,
they can be easily undone by pressing a button.
For both forms, the time required increases
with a rising number of sectors.
In \textit{V-SeMo}, geodesics do not have to be completed manually.
Thanks to the automation functions,
the necessary intermediate steps,
such as aligning the sectors and extending the lines step by step,
can be performed directly by the application.
This speeds up the process and allows
many geodesics to be constructed in a short time,
regardless of
how many sectors the model contains.

In the case of spacetimes,
the use of
virtual sector models is quite different
from the use of their paper counterparts.
Because the shape of a paper sector
cannot be easily transformed due to its physical nature,
the only way to construct geodesics by arranging sectors
side by side in the spacetime model is through
transfer sectors,
as described by \cite{kraus2019}.
In contrast, virtual sectors can be transformed
with sliders and directly connected.
This improves the clarity of the model
and simplifies the manual completion of geodesics,
reducing
the time needed to complete the geodesics
by at least half.
Utilising the automation functions
for spacetime models further simplifies
and speeds up
the construction of spacetime geodesics.

We can also analyse the transition from using paper sector models
to using the learning environment \textit{V-SeMo}
in terms of the SAMR model (\citealt{puent2006})
that describes the
roles of educational technology
in supporting learning.
In its basic mode,
\textit{V-SeMo} can be used
as a direct
alternative to paper models,
with the same manual and step-by-step methods.
This is a straightforward replacement of
analogue materials with equivalent digital representations
and corresponds to the
\textit{Substitution} level
in the SAMR model.
At the same time, \textit{V-SeMo} provides
functional improvements:
Implicit aids built into the design
support the learning process.
With these features of enhancement,
\textit{V-SeMo} is used at the \textit{Augmentation} level.
In its advanced mode, \textit{V-SeMo} provides features
that allow learning activities
impossible with pen-and-paper methods.
With the automation functions,
large models and high-resolution models become accessible.
This enables the step from basic principles to applications,
i.e. the step from the introduction of the basic concepts
using a few geodesics on a small model
to the exploration of actual spacetimes
using many geodesics and long geodesics where required.
New tasks can be formulated,
and students can work independently
and take charge of the exploration of a spacetime
when given a sector model and the tools
provided by \textit{V-SeMo}.
Also, \textit{V-SeMo} being designed as a freely available web application
allows
to make use of the opportunities associated with online
activities
for more flexible, autonomous and asynchronous learning.
Thus, with
new tasks
and new classroom strategies,
\textit{V-SeMo} also facilitates teaching
at the transformative levels
(\textit{Modification} and \textit{Redefinition}) of the SAMR model.

\subsection{Information presentation and the design of guided activities}

\begin{figure}
  \centering
  \begin{minipage}[t]{0.3\textwidth}
    \includegraphics[width=\textwidth]{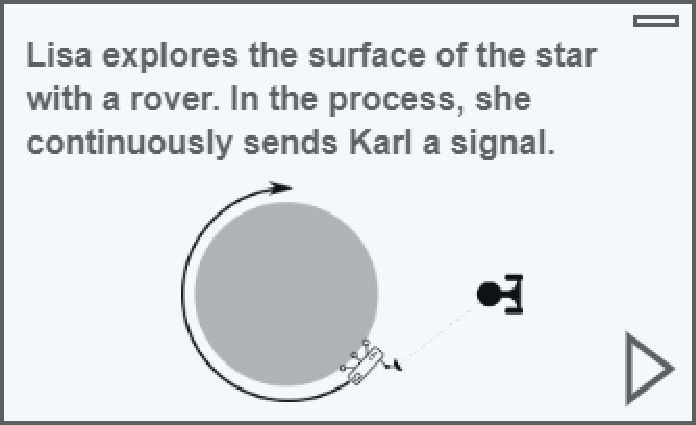}
    (a)
    \vspace{0.2cm}
  \end{minipage}
  \quad
  \begin{minipage}[t]{0.3\textwidth}
    \includegraphics[width=\textwidth]{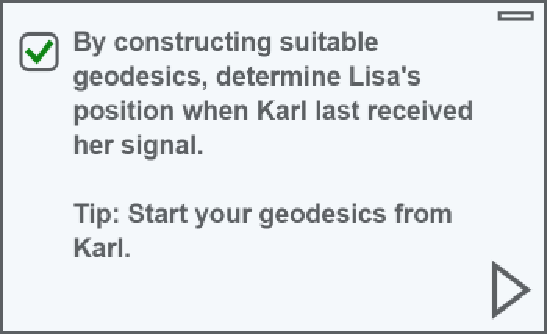}
    (b)
    \vspace{0.2cm}
  \end{minipage}
  \quad
  \begin{minipage}[t]{0.3\textwidth}
    \includegraphics[width=\textwidth]{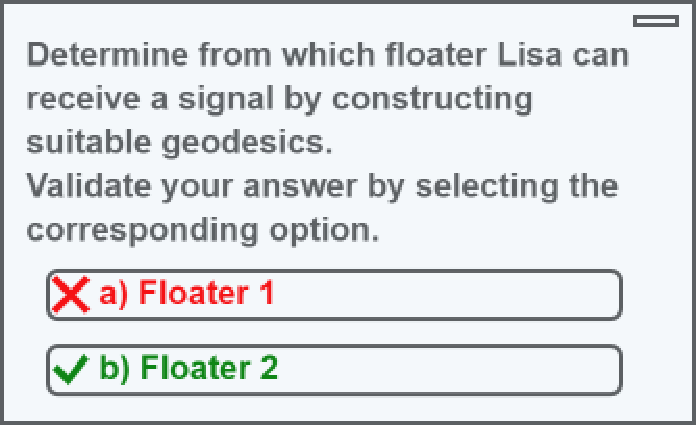}
    (c)
    \vspace{0.2cm}
  \end{minipage}
  \caption{The information panel integrates the display of text and
    images in the interface of \textit{V-SeMo} (a). It can provide
    to-do lists with checkboxes which are checked automatically after
    successful completion of the tasks (b). Also, tasks can be
    combined with questions with automatic checking of the users'
    answers (c).}%
  \label{fig:V-SeMo_exercise_boxes}
\end{figure}

To support and to structure
learners' use of \textit{V-SeMo},
we have designed an information panel
as part of the interface
(figure~\ref{fig:V-SeMo},
figure~\ref{fig:V-SeMo_exercise_boxes})\footnote{%
The creation and editing of information panels
requires adapting the code,
for access
see \cite{weiss2024}.
}.
The panel
allows to display text and images
on individual slides.
Learners switch between slides
using navigation arrows
(figure~\ref{fig:V-SeMo_exercise_boxes}(a)).
Objects in the workspace and their appearance
can be linked to individual slides.
Objects can, e.g., be visible in connection with some slides,
and hidden in connection with others,
sectors can have different positions and orientations,
geodesics can be displayed completed to varying degrees,
and the visible workspace area and zoom level
can be adjusted
ensuring that the focus is on
the relevant part of the workspace.

In addition to displaying information and images,
the information panel
facilitates the creation of to-do lists
(figure~\ref{fig:V-SeMo_exercise_boxes}(b)).
The items on the list
are associated with specific actions,
such as pressing a button,
assembling sectors,
or drawing lines over predefined points.
Each item contains a checkbox
which automatically marks the item as completed
upon successful execution of the corresponding action.
This feature creates the possibility
to break down tasks into steps
and to guide learners through these steps
while providing immediate feedback on their work.
Furthermore, to-do-lists can be combined with
questions,
and then allow automatic checking of the learners'
answers
(figure~\ref{fig:V-SeMo_exercise_boxes}(c)).
Together,
these features enable the creation of systematic guided activities,
as illustrated in the following section.
The use of ready-made guided activities
with feedback
means that learners require less instruction by teachers
and it facilitates the integration of sector model tasks
into online courses.
It thus opens up new ways of teaching,
further supporting teaching at the transformative levels
of the SAMR model.

%----------------------------------------------------------

\section{Guided activity ``Journey to a Neutron Star''}
\label{sec:teaching_example}

Here we present a learner activity
that illustrates the potential of \textit{V-SeMo}.
In this unit learners use \textit{V-SeMo}
to investigate light deflection
in the vicinity of a neutron star.
The unit is in the form of a guided activity.
Before embarking on this unit,
learners should be familiar with
the concept of geodesics and
with the fact that light rays are geodesics.
They should also be familiar
with the concept of sector models,
the construction of geodesics on sector models,
and the operation of \textit{V-SeMo}
including the `Manipulate Geodesics' mode.
The guided activity is accessible
in the online resources for this article
(\citealt{weiss2024}).

\subsection{The guided activity}

\begin{figure}
  \centering
  \includegraphics[width=\textwidth]{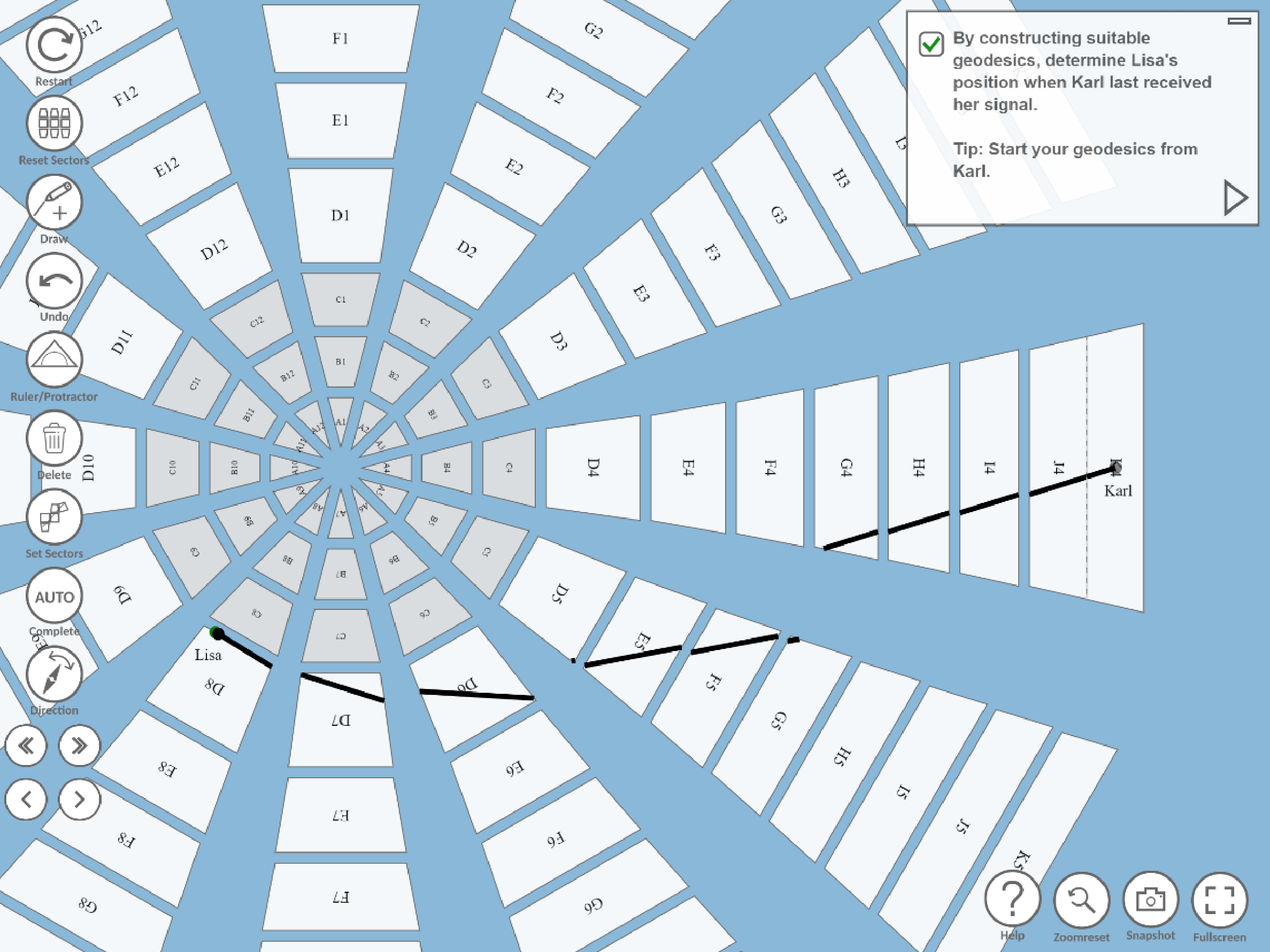}
  \caption{Screenshot of \textit{V-SeMo} in the activity ``Journey to
    a Neutron Star''. The sector model represents the equatorial plane
    both inside the neutron star (grey sectors) and outside the star
    (white sectors). The information panel sets a task (construction
    of a certain geodesic), and the ticked checkbox signals successful
    completion.}%
  \label{fig:v-semo_neutronstar}
\end{figure}

``Journey to a Neutron Star'' lets learners explore
the Schwarzschild spacetime exterior to
a non-rotating, spherically symmetric neutron star.
It makes use of
the sector model shown in
figure~\ref{fig:v-semo_neutronstar}
(described in more detail below)
that represents the equatorial plane,
i.e. a symmetry plane of the three-dimensional space
at constant Schwarzschild time.
The information panel
is used to tell the story of two astronauts
who travel to the neutron star.
The slides provide several tasks
embedded in this story
and give feedback upon their completion.
The tasks require learners
to construct geodesics in the equatorial plane
and study their properties.
Since the geodesics
on this sector model
are purely spatial geodesics,
they are in the nature of an analogy
to the spacetime null geodesics
that represent light rays.
We further comment on this analogy below.

Table~\ref{tab: guidedactivity}
displays the content of the information panel
that structures this guided activity.
Slides~1 and~2 set the scene:
Astronauts Karl and Lisa
arrive at a neutron star,
Karl remains on board the base ship,
and Lisa lands on the stellar surface.
Slides~3 and~4 set the first task:
Lisa wants to communicate the successful
landing by sending a light signal
to Karl.
\textit{V-SeMo} displays a geodesic starting
at Lisa's position, but not reaching Karl.
The learner is asked to make the geodesic
reach Karl
by adjusting its starting direction.
This is effected using
the `Direction' function
in `Manipulate Geodesics' mode
to systematically change
the starting direction.
Slide~4 contains a to-do-list,
and the
correct choice
of starting direction
prompts a confirmation
and allows the learner to proceed to the next slide.
Slides~5 and~6 continue the story:
Lisa explores the surface and
continuously sends signals to Karl.
\textit{V-SeMo} displays Lisa's positions
and the respective geodesics.
When Lisa moves
into the far side
of the neutron star,
at some point, her signal does not reach Karl any more.
Slide~7 again sets a task:
The learner is asked to find the last position
from which Lisa's signal can still reach Karl.
This can be accomplished
by drawing a geodesic
starting at the position of Karl,
and then modifying its starting direction.
Again, a to-do-list signals
the completion of the task.
The result is displayed in
figure~\ref{fig:v-semo_neutronstar}:
Karl receives a signal
from the far side of the star.
As a result of this activity,
learners discover that
observers can in part see
the side of a neutron star
that is facing away from them.

\begin{figure}
  \centering
  \includegraphics[width=0.75\textwidth]{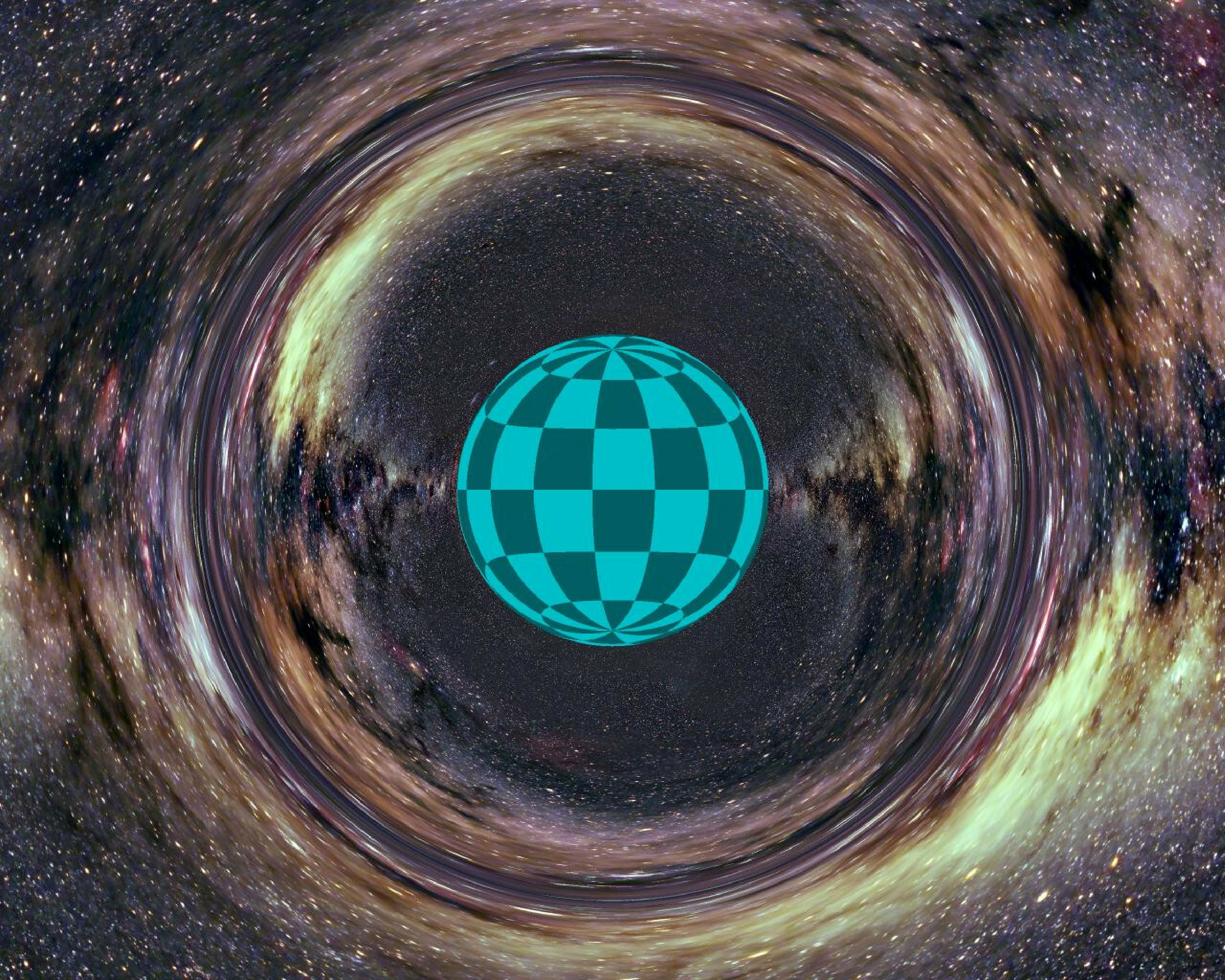}
  \caption{Simulated view of a neutron star in front of the Milky Way
    taking light deflection into account. For the purpose of
    illustration, the stellar surface is provided with a chequered
    pattern. The surface of the star is located at 2~Schwarzschild
    radii, and the virtual camera is positioned at a radial coordinate
    of 20 Schwarzschild radii. Image: Thomas Reiber.}%
  \label{fig:neutronstar_lightbending}
\end{figure}

With figure~\ref{fig:neutronstar_lightbending}
one can further illustrate this result and show
what a neutron star would look like
for a nearby observer.
The computer simulated image
depicts a neutron star in front of the Milky Way,
artificially providing the neutron star surface
with a chequered pattern
that helps to
show up geometric effects.
Each check has an extent of 30 degrees in length
and 30 degrees in width.
One can clearly
see that the visible part of the surface
extends beyond both poles.
This can also be seen along the equator:
There are about eight checks visible here,
not six as would be expected.
The Milky Way in the background
appears distorted by the deflection of its light
near the neutron star.

Finally, one should add that
the story of Karl and Lisa
is clearly quite unrealistic.
Because of the large interstellar distances,
no human being can reach a neutron star.
And even if they did, they could not safely
land on it:
The gravitational acceleration
on the surface of a neutron star is about 200 billion times
that on Earth.

The tasks in ``Journey to a Neutron Star''
illustrate how
the virtual representation of sector models
enhances their possible applications:
Both the large number of sectors needed for this activity
and the large number of geodesics to be constructed
are manageable
thanks to the tools for geodesic manipulation.
The work involved
in carrying out these tasks
without the advanced features or
with paper sector models would
be prohibitive.

  \begin{table}[htbp!]
    \caption{\label{tab: guidedactivity}
    Content of the information panel for the
    guided activity ``Journey to a Neutron Star''.
    }

    \begin{tabular}{p{9.5cm}p{5cm}}
      \br
      \textbf{
      Slides of the information panel
      }
      &
      \textbf{
      Learner actions
      }\\
      \mr
      \begin{itemize}
        \item[S1:] Karl and Lisa arrive in sector K4 near the neutron star.
        \item[S2:] Lisa leaves the spaceship with a lander
          and lands in sector D5 on the surface of the star.
      \end{itemize}
      & \vspace{-0.2cm}
      \includegraphics[width=1\linewidth]{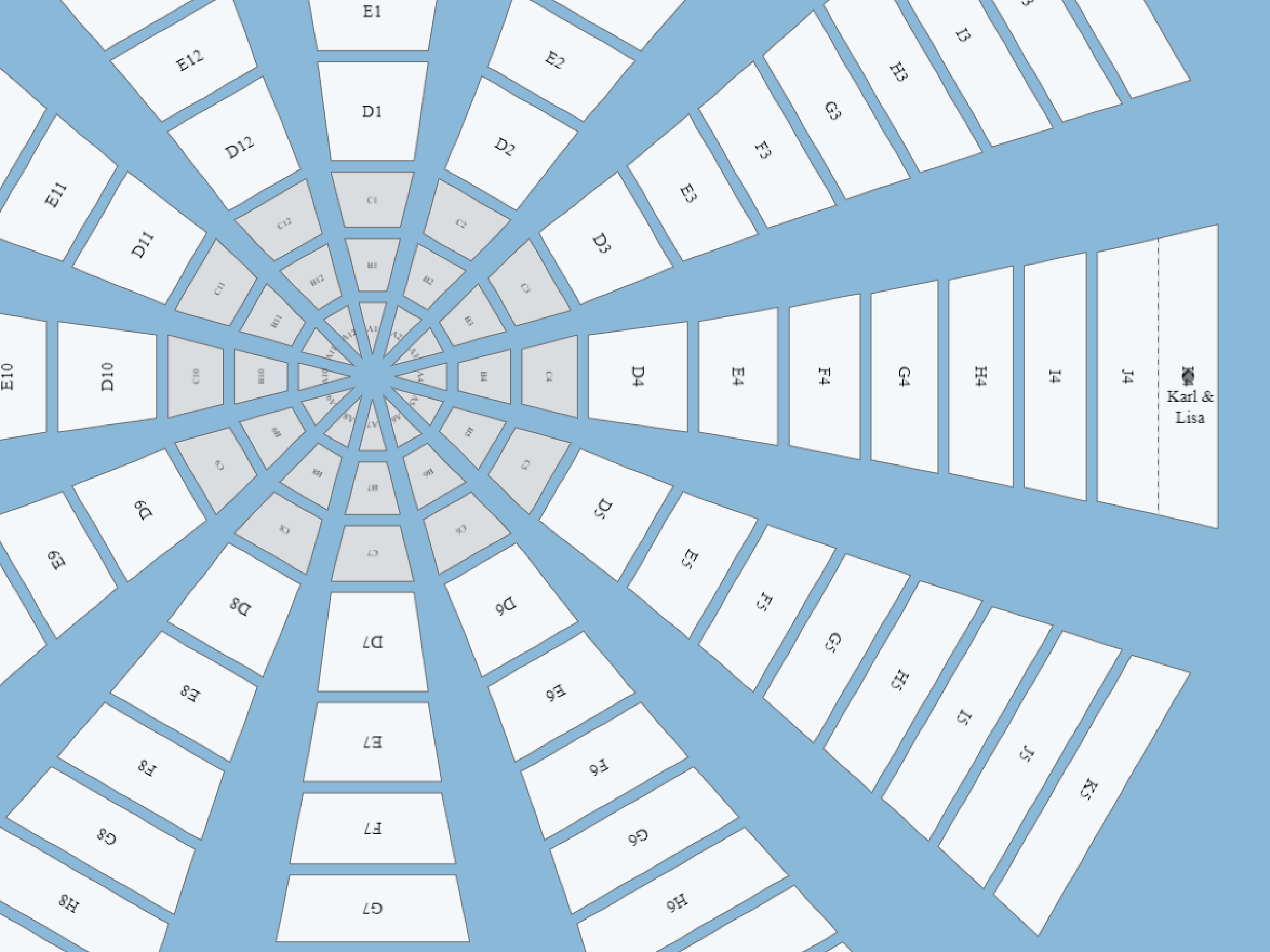}\\
      \mr
      \begin{itemize}
        \item[S3:] To tell Karl about her successful landing,
        Lisa wants to send him a signal (blue geodesic).
        \item[S4:] Help Lisa orientate her transmitter correctly.
          First select the blue geodesic.
          Use the `Direction'
          button to change its starting direction.
      \end{itemize}
      & \vspace{-0.2cm}
      \includegraphics[width=1\linewidth]{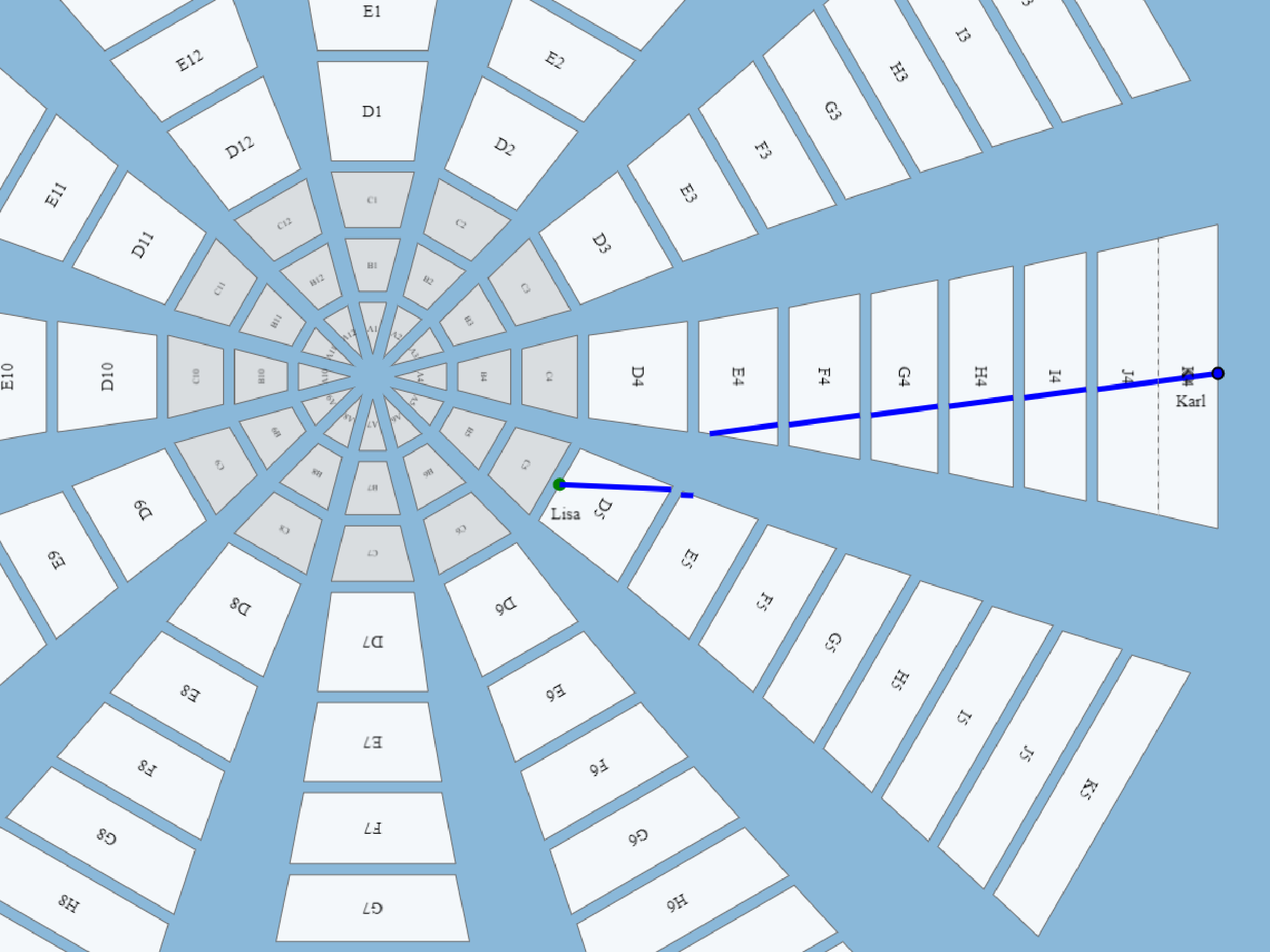}\\
      \mr
      \begin{itemize}
        \item[S5:] Lisa explores the surface of the star with a rover.
        In the process, she continuously sends signals to Karl.
      \end{itemize}
      & \vspace{-0.2cm}
      \includegraphics[width=1\linewidth]{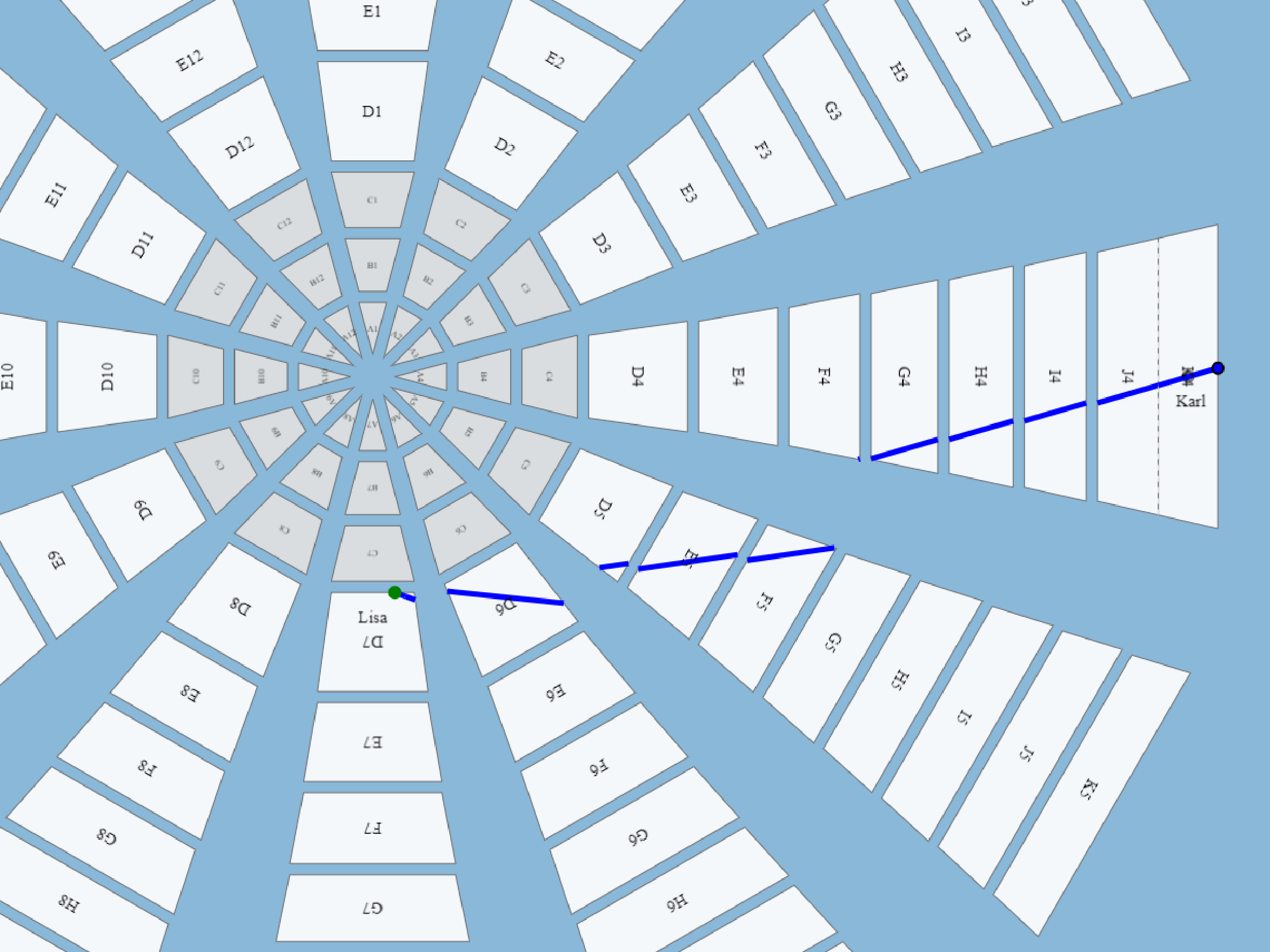}\\
      \mr
      \begin{itemize}
        \item[S6:] When Lisa reaches a certain position,
        Karl can no longer receive her signal.
        \item[S7:] By constructing a suitable geodesic,
          determine Lisa's position
          when Karl last received her signal.
          Tip: Start your geodesics
          from the position of Karl.
        \end{itemize}
      & \vspace{-0.2cm}
      \includegraphics[width=1\linewidth]{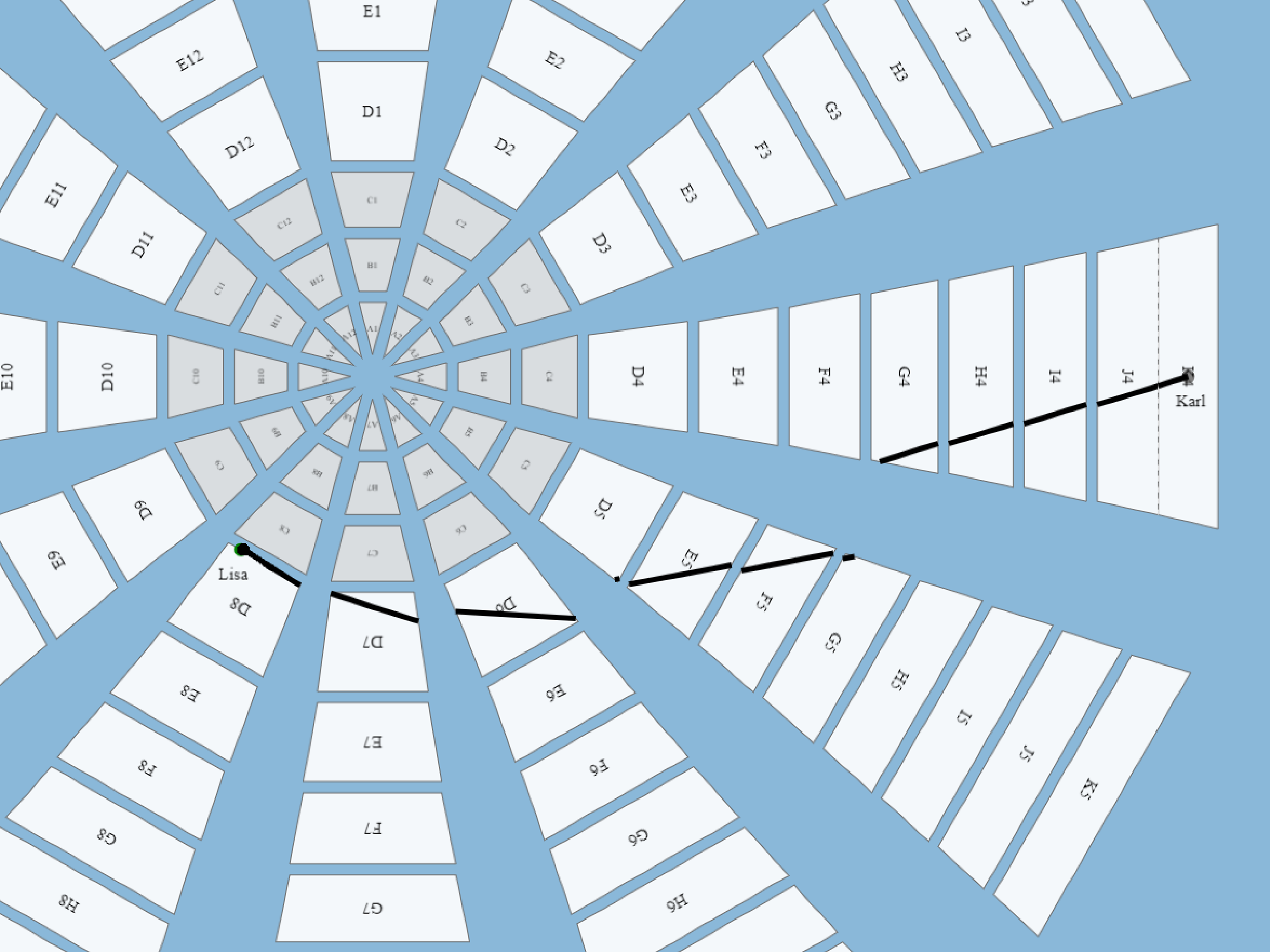}\\
      \br
    \end{tabular}
  \end{table}

\subsection{The sector model for ``Journey to a Neutron Star''}

For this activity we assume a non-rotating,
spherically symmetric neutron star.
The interior of the star
(that plays no role in this activity)
is described by a simple homogeneous density model
(\citealt{shap1983}).
Outside the star,
the spacetime is given
by the Schwarzschild metric.
Restricted to the equatorial plane, it reads

\begin{eqnarray}
  \mathrm{d}s^{2} = \frac{1}{1-\rs/r}\mathrm{d}r^{2} + r^{2}\mathrm{d}\phi ^{2}
  \label{eqn:schwarzschildmetric}
\end{eqnarray}

with the usual Schwarzschild radial coordinate $r$
and azimuthal angle $\phi$.
Here, $\rs = 2GM/c^2$ is the Schwarzschild radius
of the neutron star with mass $M$,
$G$ is the Newtonian gravitational
constant and $c$ the speed of light.
In the activity ``Journey to a Neutron Star'',
geodesics are constructed in the equatorial plane.
These are spacelike geodesics
that we use in analogy to the spacetime null geodesics
of light signals.
This is a well-established analogy
which has often been used together with
embedding surfaces of the equatorial plane
(e.g. \citealt{dinv1992}).
The deflection of these spatial geodesics, however,
is smaller than that of spacetime null geodesics
starting in the same spatial direction.
For the activity, we would like the deflection
and thereby the physical phenomena
to be the right order of magnitude, and
we obtain this by setting
the neutron star parameters accordingly.
For light deflected by a neutron star
with a `canonical' mass of 1.4 solar masses
and a radius of 8.5 km (a fairly compact star,
which we choose for a large light deflection effect),
the maximum deflection angle
for a path from the neutron star surface to infinity
is 58 degrees,
obtained for a path that grazes the surface of the star.
We obtain the same maximum deflection angle
for the spatial geodesics in the equatorial plane
by assigning 1.4 solar masses and 5 km ($1.2\rs$) radius
to the neutron star.

The sector model
represents the equatorial plane
for radial coordinate $r$
between~$r=0$ and $r=4.4\rs$,
with the surface of the star
at $1.2\rs$
for the reasons described above.
This part of the equatorial plane
is subdivided into elements of area
that are quadrilaterals
with vertices at
$30^{\circ}$-intervals in azimuthal angle $\phi$
and intervals of $0.4\rs$ in radial coordinate $r$.
The sectors are constructed as flat, symmetric trapezia
with the same edge lengths as the respective elements of area
(for details on the construction of sector models, see \citealt{zahn2019}).

\section{Evaluation}
\label{sec:learning_effectivness}

We report on an evaluation
in which we investigate learning results
on the concept of geodesics
for
a sector model-based teaching unit
on geodesics and light deflection.
As part of this evaluation
we also compare learning results between two groups,
one of which worked exclusively with \textit{V-SeMo}
while the other
used paper models in the introductory part
and switched to \textit{V-SeMo} for the more advanced tasks.
(We did not work with
a completely paper-based implementation
because the more advanced tasks
would have been
tedious.)

The participants are 88 secondary school students
in grades~11 and~12 (ages 16 to 17).
They are the members of~6 physics classes
who visited the University of Hildesheim
to take part in 90-minute
workshops on relativity.
All participants
(or their parent or legal guardian
in the case of children under 16)
gave written informed consent
to participate in the study.
The students had not previously studied general relativity,
and they had not
encountered non-Euclidean geometry.

For the purpose of comparison,
four of the six classes were split up
into two groups each,
one of which started the unit with paper sector models.
Splitting
classes rather than assigning whole classes
to one or to the other procedure
should ensure
that we compare groups
with the same previous knowledge
and physics background.
In total, 32 students worked with paper
sector models initially,
while 30 of their classmates
worked with \textit{V-SeMo} throughout
as well as the 26 members
of the other two classes.
Data were collected during and at the end
of the sessions. Here, we report on the
analysis of the final task,
an open question probing
students' understanding of geodesics
and light deflection.
The question was answered individually
and in written form.
For the analysis,
the answers were categorised
according to
levels of understanding
defined with reference to
the intended learning outcome.

\begin{figure}
  \centering
  \begin{minipage}[t]{0.48\textwidth}
    \includegraphics[width=\textwidth]{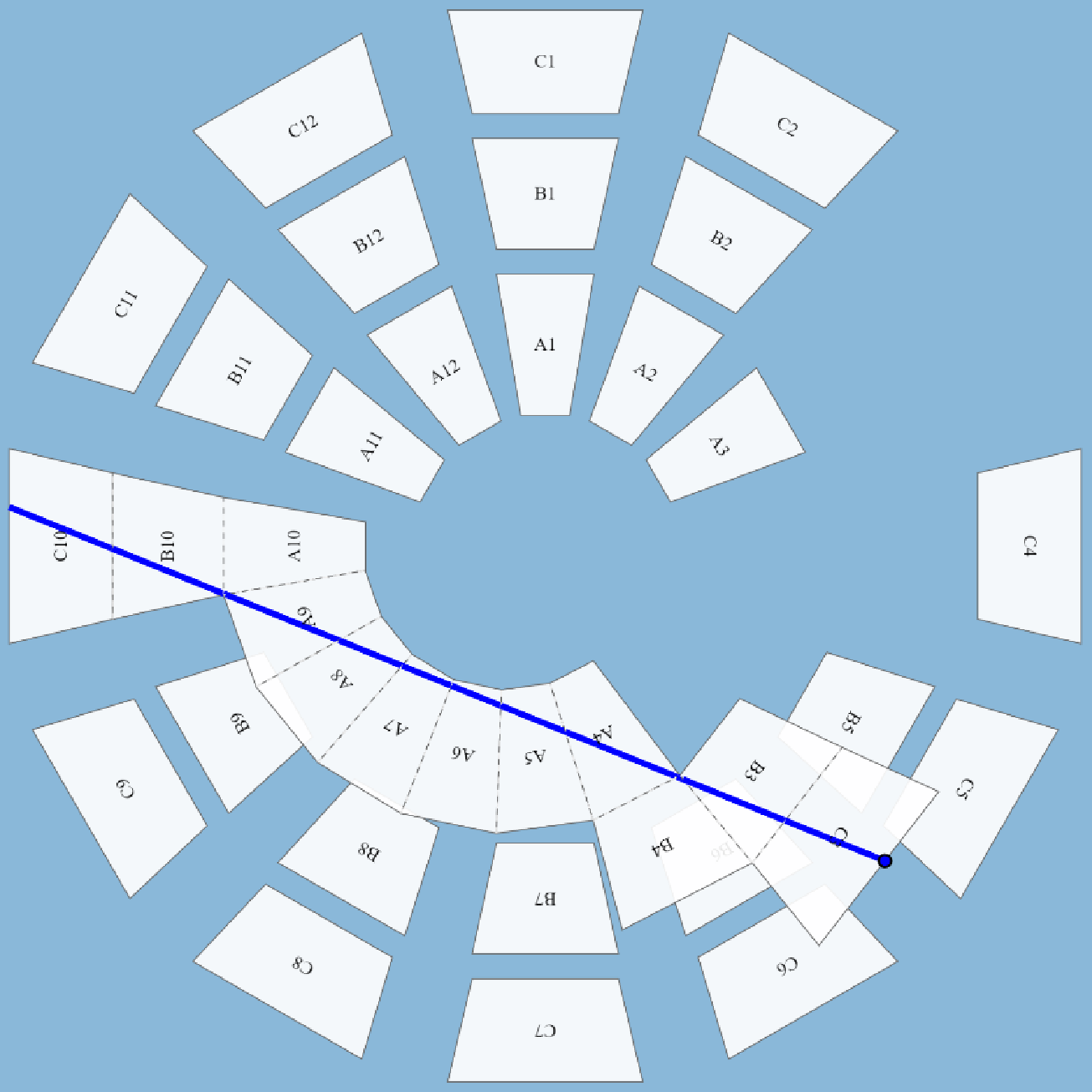}
    (a)
    \vspace{0.2cm}
  \end{minipage}
  \hfil
  \begin{minipage}[t]{0.48\textwidth}
    \includegraphics[width=\textwidth]{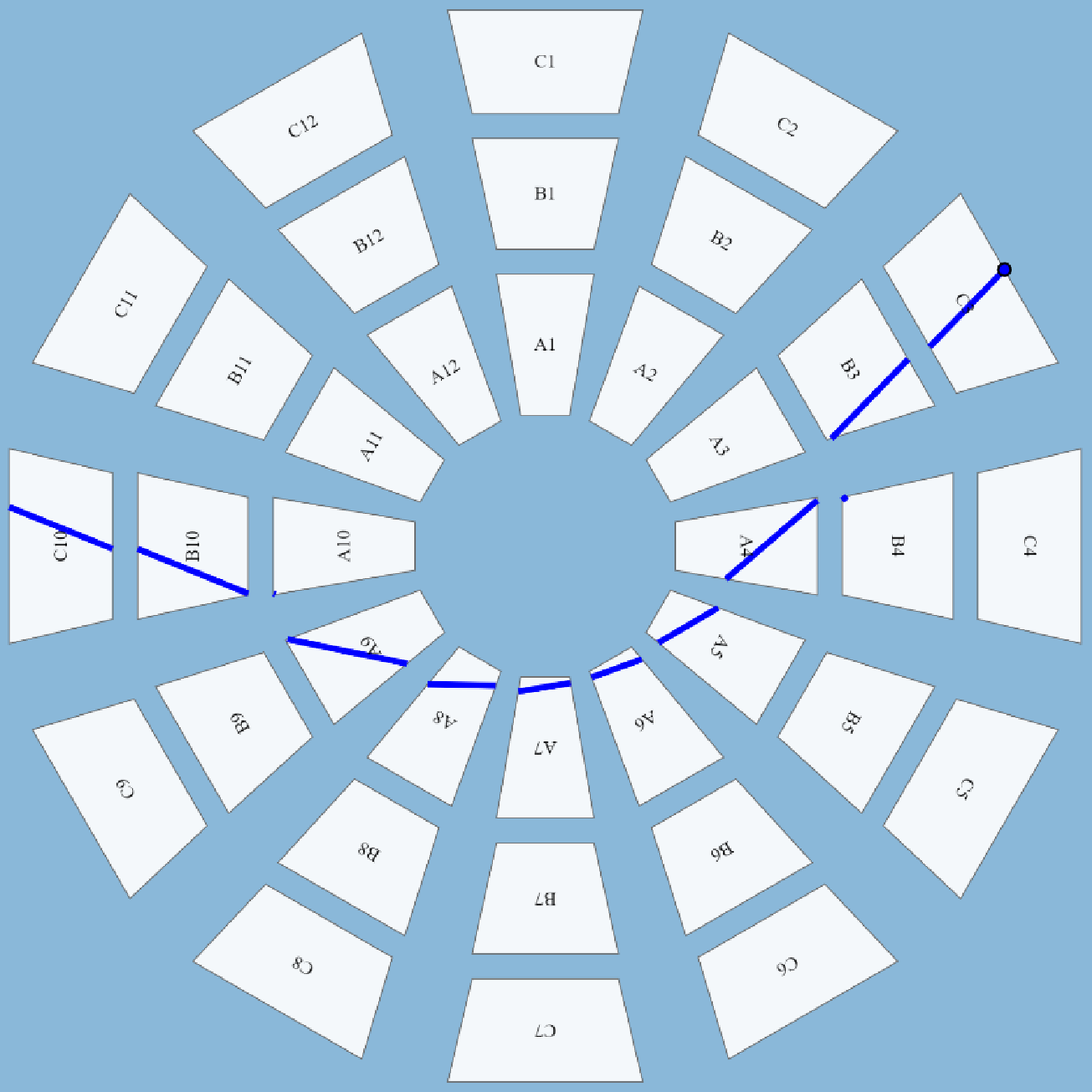}
    (b)
    \vspace{0.2cm}
  \end{minipage}
  \begin{minipage}[t]{0.9\textwidth}
  \includegraphics[width=\textwidth]{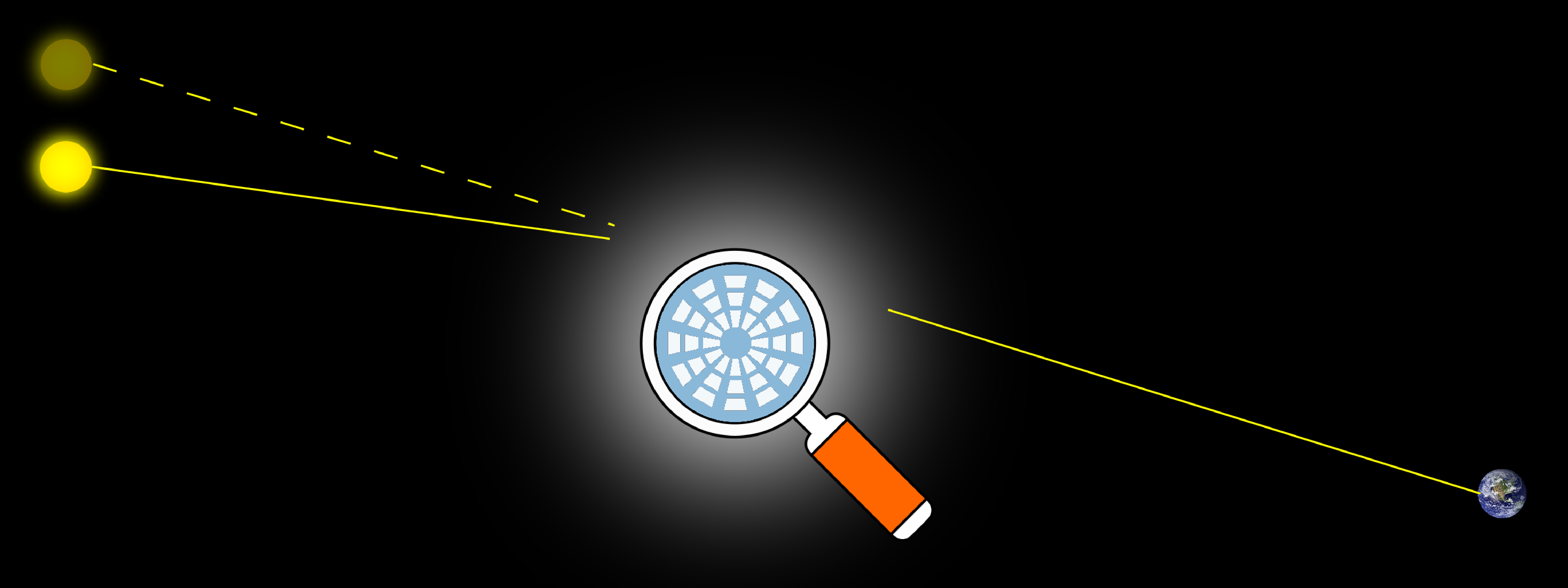}
    (c)
    \end{minipage}
  \caption{Geodesics and light deflection.
     On the sector model of the equatorial plane
     of a black hole,
     a spacelike geodesic is constructed
     as an analogy to a lightlike geodesic in spacetime.
     (a) Learners draw a geodesic,
     i.e., a locally straight line,
     that passes close to the black hole.
     (b) The sectors are rearranged
     in symmetric configuration to show that,
     globally,
     the directions before and after
     passing close to the black hole differ.
     (c) This is observed as gravitational
     light deflection.}%
  \label{fig:V-SeMo_task}
\end{figure}

The participants worked through the unit
``Geodesics and Light Deflection''
(\citealt{kraus2021}, \citealt{weiss2022})
that introduces geodesics,
lets learners construct geodesics
on several sector models,
and discusses phenomena of
gravitational light deflection.
In
the introductory part of this unit,
students work out
how to construct geodesics
on sector models,
and thereby perform the
step-by-step construction
of
geodesics on the sector model
of the sphere
(similar to figure~\ref{fig:sectors_in_action}(a)).
In the case of the four classes
that were split into
two groups each,
one of the respective groups
used paper sector models, pens and rulers
at this point
for their first go at constructing geodesics;
the other group used \textit{V-SeMo} from the start.
After this exercise,
the ``paper group''
switched to the use of \textit{V-SeMo}
and both groups continued
with identical tasks.
A key task with respect to the
final question on which we report here
is shown in figure~\ref{fig:V-SeMo_task}.
Students use a sector model
that represents the equatorial plane
not far outside the horizon of a black hole.
At this point, they have heard
that a geodesic, by definition,
is a locally straight line and
that light rays are geodesics.
In this task they construct a light ray passing close to the black
hole by drawing a straight line on the sector model
(figure~\ref{fig:V-SeMo_task}(a)).
The sectors are then rearranged in symmetric form
to show
that, globally,
the directions of the light ray
before and after passing close to the black
hole differ (figure~\ref{fig:V-SeMo_task}(b)).
This illustrates how light
always runs straight locally and
may still change its direction globally
by traversing a region of curved space
(figure~\ref{fig:V-SeMo_task}(c)).

The intended learning outcome
tested in the final question
is an understanding
of geodesics and light deflection
that includes the following points:
\begin{enumerate}
\item[1.] A geodesic is a locally straight line,
i.e. a ray of light, being a geodesic, is straight.
\item[2.] A geodesic passing close to a black hole
(or other massive object)
has a different direction far behind the object
than far in front of it,
i.e. light is deflected by such an object.
There is no contradiction with point 1.
\item[3.] The geometry of space (actually spacetime)
near the black hole
is the cause of light deflection.
\end{enumerate}

\begin{figure}
\centering
\fbox{\includegraphics[width=0.75\textwidth]{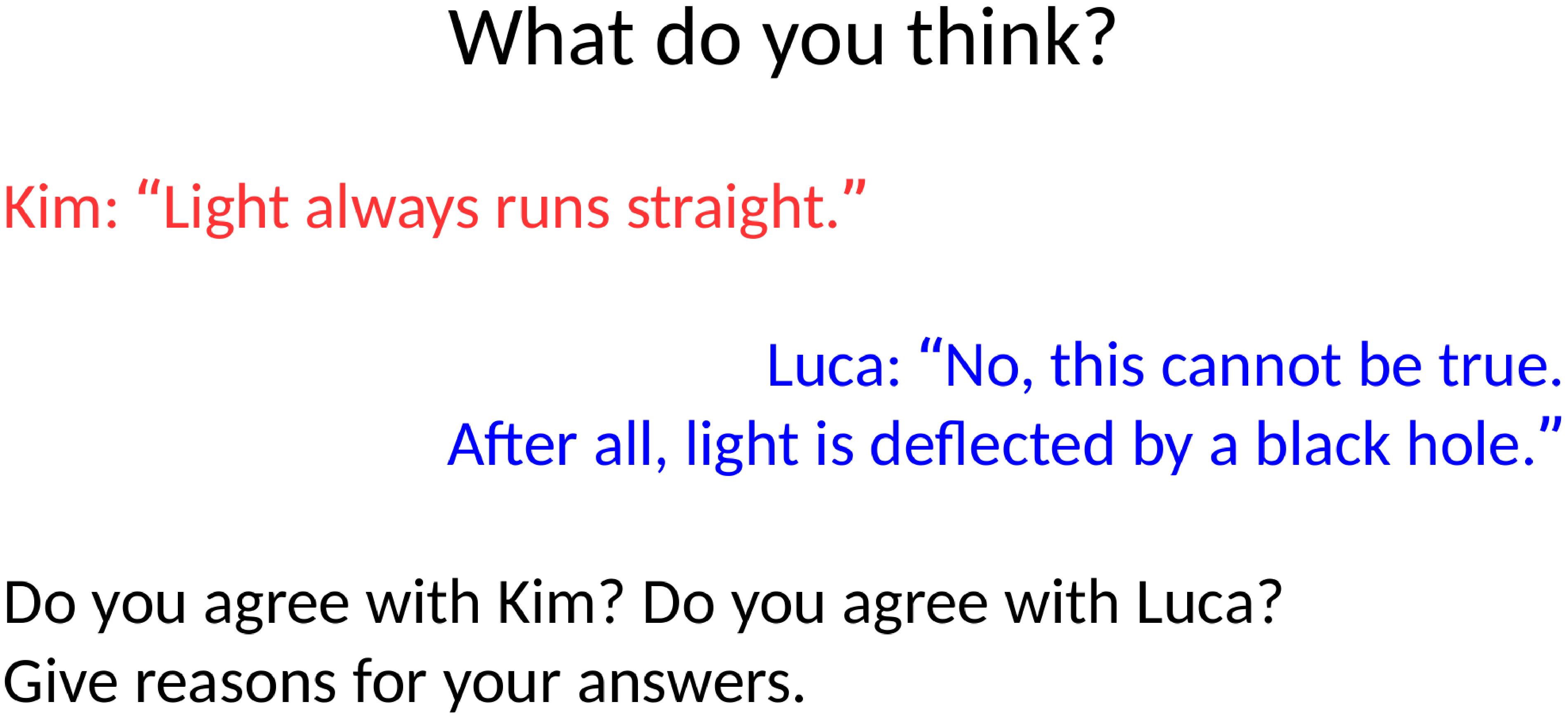}}
\caption{Hand-out with the final question in the teaching unit,
  prompting students to express their ideas about geodesics and light
  deflection.}%
\label{fig:concept_cartoon_en}
\end{figure}

For the final task,
a fictitious exchange of opinions
was used to elicit students' ideas about geodesics and
light deflection.
Students were asked
to take a stand on two statements
(as shown in figure~\ref{fig:concept_cartoon_en})
and to justify their position,
providing an answer in writing.

We consider this to be a
fairly
challenging task.
Firstly, because
the task is not about reproducing a fact,
but about discussing a connection
between two aspects of the unit.
These
are juxtaposed
by proposing an apparent contradiction.
Students have to
link their work on the sector models
(drawing geodesics as straight lines)
with the astronomical implications.
Secondly, students have to express their
understanding in their own words.
Non-euclidean geometry being a new topic
for them,
this is a high linguistic demand, too.
We therefore expect the answers
to cover a wide range
of levels in sophistication.

The responses of the 88 participants
were categorised into three levels
of understanding:
``advanced'', ``basic'', and ``sub-basic'' and
the fourth category ``non analysable''
(table~\ref{tab: kim_luca_overview}).
First,
the answers that were deemed incomprehensible
or inconclusive,
a total of 10 answers,
were assigned to
the category ``non-analysable''
and were excluded
from further analysis.
The remaining 78 responses were then analysed in two steps.
Step one considers
understanding on a basic level.
If a student expresses point 1 above,
e.g., by stating that a light ray is a geodesic,
or that a light ray is straight,
the response is categorised as
expressing basic understanding.
Responses that
reject the idea of light rays being straight
and responses that
are contradictory in themselves
or express other misconceptions
concerning light propagation,
are categorised as
expressing sub-basic understanding.
In step two,
the responses expressing basic understanding
were reexamined
with a view to point 2 above.
When a student agrees that there is
light deflection
and expresses the view that this is no
contradiction
to light rays being straight or being geodesics,
the response is re-categorised as
advanced level.
Otherwise, it is basic level.

\begin{table}[htbp!]
  \centering
    \caption{\label{tab: kim_luca_overview}%
    Answers to the final question categorised into four levels.
    }
    \begin{tabular}{lllp{10.cm}}
      \br
      \textbf{category} & \textbf{n} & \textbf{\%} & \textbf{example} \\
      \mr
      \makecell{advanced level} & 44 & 50\;\% &
      ``I agree with both because light always follows geodesics.
      These always run straight. Nevertheless they can be deflected by
      black holes.'' \\
      \mr
      basic level & 10 & 11\;\% &
      ``In my opinion Kim is right
      because light always travels in a geodesic,
      like a straight line.'' \\
      \mr
      sub-basic level & 24 & 28\;\% &
      ``By itself, light always travels in a straight line.
      However, the black hole deflects the light in such a way,
      that it is no longer straight. Both are right.'' \\
   \mr
      not analysable & 10 & 11\;\% &
      ``I agree with Kim,
      but Luca's approach is correct.
      The light appears to bend
      due to the effect of the black hole.
      However, this image is only in theory
      and not in reality.'' \\
      \br
    \end{tabular}
  \end{table}

\begin{figure}[htbp!]
  \centering
  \includegraphics[width=1\textwidth]{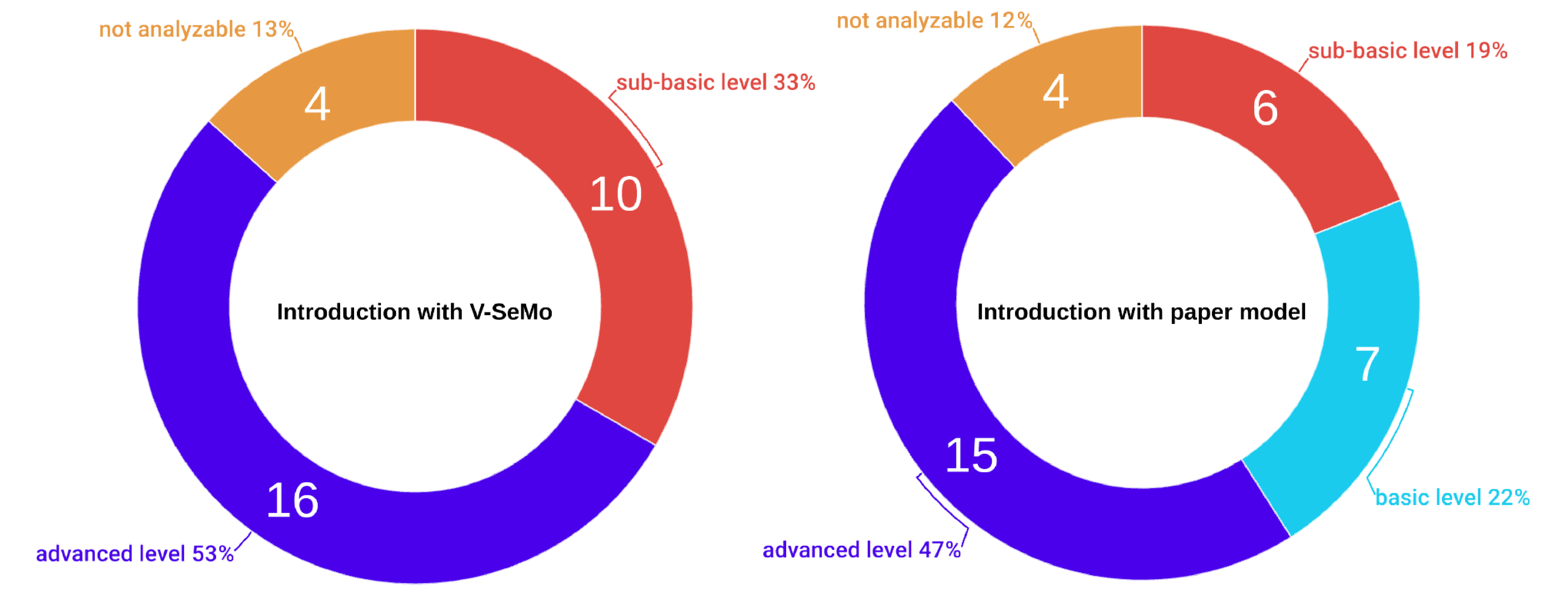}
  \caption{Comparison of the ``\textit{V-SeMo}-only'' and ``paper''
    groups. The pie charts show the categorisation of the answers into
    the four levels listed in table~\protect\ref{tab:
      kim_luca_overview}. The total number is 30 for the
    ``\textit{V-SeMo}-only'' group (left) and 32 for the ``paper''
    group (right).}%
    \label{fig:pie_charts}
\end{figure}

Table~\ref{tab: kim_luca_overview} summarises the results
of this analysis,
and also provides typical examples for answers
in the four categories.
Half of the participants (44) gave an advanced level answer,
10 answers were basic level,
24 answers were sub-basic level,
and 10 answers were not analysable.
Given the high level of difficulty of this task
as described above
we consider the rate of about 50\;\% advanced level answers
as quite satisfactory.
In 33 of the 54 basic or advanced level responses,
there are, additionally, references to point 3 above.
These go beyond the expected answer to this final question
and attribute the run of the geodesics
to the geometric properties of the traversed space.
In figure~\ref{fig:pie_charts}
we compare the results of the
``paper'' group and the
``\textit{V-SeMo}-only'' group.
The pie charts
show the
categorisation of the answers
for the two groups
separately.
In both cases
half of the responses
are advanced level.
Thus,
the fraction of 50\;\% of the students
reaching the intended learning outcome
is obtained both with and without
the use of paper sector models
in the introductory part,
establishing
comparable learning effectiveness
for both types of sector representation.
Answers at basic level
account for about 20\;\% in the ``paper'' group,
and are absent in the ``\textit{V-SeMo}-only'' group.
This may be an indication
that some students profited
from the hands-on experience
of working with material artefacts,
pen and ruler.
Given the small absolute number
of students answers at this level,
further tests
are needed to investigate
this potential difference between
the two groups.

The causes for student answers
on the basic and lower levels
remain to be studied in detail.
Our experience from workshops suggests
that the
reasons
for imperfect answers
include
language and misconceptions.
Language in the sense that,
in this response format,
incorrect wording or
unclear expressions
may hide the fact
that a learner has indeed developed correct ideas.
Also, we have occasionally noticed
that individual students
have misconceptions that appear to be
based on popular science accounts.
This is quite conceivable in the present case
since descriptions of gravitational light deflection
typically refer to light rays as bent or curved,
but do not as a rule stress
the fundamental principle that,
in spacetime,
light follows locally straight lines.
It
may also play a role that
the workshops that we evaluated
were not part of the participants'
regular physics classes
and therefore not relevant for
student assessment.
Finally, the evaluation concerns a single
90-minute workshop. Inclusion of this unit
in a longer teaching sequence
with time for repetition and practice
should be expected to increase the fraction of
participants
developing advanced level understanding.
Further tests may probe the relevance of these ideas.

%--------------------------------------------------

\section{Discussion and conclusions}
\label{sec:discussion}

We have reported on the development
and evaluation
of the learning environment
\textit{V-SeMo}.
\textit{V-SeMo} presents
a virtual representation of two-dimensional sector models
and provides tools for their handling,
facilitating the study of curved surfaces, spaces, and spacetimes.
We have pointed out
the enhanced capabilities of
virtual sector models
compared with paper models
including automation features
that
make large models and a large number of geodesics accessible,
features that support individual learning,
and online capabilities that facilitate flexible classroom use.

\textit{V-SeMo} has been tested extensively:
At the technical level
for computer and tablet use
with different operating systems and browsers,
at the design stage
with specially conducted test series,
and in its current advanced stage of development
with a study that
confirms the learning effectiveness of a \textit{V-SeMo}-based course
on light deflection.
In total, more than 650
learners
have worked with \textit{V-SeMo}
in individual interview sessions or group workshops.

With the development of \textit{V-SeMo}
we aim to provide new
learner activities,
in particular
for those learners of general relativity
who follow courses
with mathematics restricted to secondary school level.
At the present time,
the focus is
on
the Schwarzschild spacetime,
an example that is both comparatively simple
and highly relevant,
and in fact is probably the spacetime that is
most extensively treated in textbooks.
The sector model approach, however,
is applicable to other spacetimes as well,
and so is the set of tools provided by \textit{V-SeMo}.
Thus,
the development of \textit{V-SeMo} has created a basis
for
a broad range of prospective
learner activities in general relativity.

%----------------------------------------------------------

\section*{ORCID iDs}

S Weissenborn 0000-0002-3537-768X \\
U Kraus 0000-0002-0549-5255 \\
C Zahn 0000-0001-5348-6698

%--------------------------------------------------

\newcommand{\newblock}{}

\end{document}